\documentclass[aps,prc,twocolumn,superscriptaddress,floatfix]{revtex4-2}

\RequirePackage{latexsym,amsmath,amssymb,amsfonts}
\RequirePackage{multirow}
\RequirePackage{graphicx}
\RequirePackage[colorlinks,citecolor=blue,urlcolor=blue,linkcolor=blue]{hyperref}

\begin{document}

%Title of paper
\title{Parameters of nucleon densities and the Coulomb barrier in heavy-ion collisions}

\author{Makar Simonov}
\email[Corresponding author: ]{makar.simonov@exp2.physik.uni-giessen.de}
\affiliation{II. Physikalisches Institut, Justus-Liebig-Universit\"at Gie\ss en, 35392 Gie\ss en, Germany}
\affiliation{Joint Institute for Nuclear Research, 141980 Dubna, Russia}
\affiliation{Lomonosov Moscow State University, 119991 Moscow, Russia}

\author{Alexander Karpov}
\affiliation{Joint Institute for Nuclear Research, 141980 Dubna, Russia}
\affiliation{Dubna State University, 141982 Dubna, Russia}

\author{Tatiana Tretyakova}
\affiliation{Joint Institute for Nuclear Research, 141980 Dubna, Russia}
\affiliation{Lomonosov Moscow State University, 119991 Moscow, Russia}

\date{\today}

\begin{abstract}
\begin{description}
\item[Background] When modeling nuclear processes which occur in heavy-ion reactions, it is necessary to calculate the potential energy of interaction between two nuclei. One of the main features determining the dynamics of the nucleus-nucleus collision is the Coulomb barrier, the knowledge of which is especially important when low- and intermediate-energy reactions are being studied.
\item[Purpose] Our goal is to establish a parameterization of nucleon density distributions for calculation of the nucleus-nucleus double-folding potential in nuclear reactions. Special attention is paid to the description of the Coulomb barrier.
\item[Method] The study analyzes experimental data on charge radii, diffuseness, and neutron skin thickness of atomic nuclei. The nucleus-nucleus potential is calculated in the framework of the double-folding method with the effective nucleon-nucleon interaction taken in the form of the zero-range Migdal potential. Based on this analysis and comparison with the Bass potential, parameters of nucleon density distributions are fitted to reproduce the Coulomb barrier.
\item[Results] A method for correcting the parameters of nucleon densities to reproduce the Coulomb barrier with the double-folding potential is proposed. 
\item[Conclusions] The presented way to correct nucleon densities allows obtaining a satisfactory description of the Coulomb barrier that is important for modeling near-barrier collisions of heavy ions.
\end{description}
\end{abstract}

\maketitle

\section{Introduction \label{intro}}
    Synthesis of heavy exotic nuclei is one of the most intriguing research tasks of today in fundamental nuclear physics. Study of new nuclides allows to broaden our understanding of properties of rare and insufficiently explored isotopes, limits of nuclear stability, and mechanisms of nuclear reactions~\cite{Giuliani2019}.
    
    The heaviest isotopes of high atomic number elements, especially superheavy elements, contain a high number of neutrons. Increased neutron-to-proton ratio in comparison with $\beta$-stable isotopes hinders the synthesis of new heavy nuclei. Cold and hot fusion methods in low-energy reactions with heavy ions, in particular ${}^{48}$Ca, have made it possible to obtain a wide range of new nuclides up to ${}^{294}$Og~\cite{Hofmann2015,Oganessian2015}. However, attempts to synthesize isotopes of elements 119--120 have been unsuccessful so far~\cite{Oganessian2017,haba2019}, and techniques to produce nuclides lying on the ``island of stability'' around $Z \sim 114$, $N \sim 184$ are being studied~\cite{Ackermann2017}. At present, it is proposed to use fusion reactions between stable ions such as ${}^{50}$Ti, ${}^{51}$V, ${}^{54}$Cr and actinides with $Z$ = 94--98~\cite{Sridhar2015,Voinov2020,kayumov2022}. Multinucleon transfer reactions involving stable and long-lived isotopes, e.g. ${}^{198}$Pt, ${}^{208}$Pb, ${}^{232}$Th,  ${}^{238}$U,  ${}^{251}$Cf, are considered to be a promising way to obtain heavy nuclides in a wide range of masses~\cite{Tian2008,Watanabe2015,Saiko2022}. The possibility of forming a new nucleus in nuclear reactions is tightly connected to the peculiarities of nucleus-nucleus interaction.
    
    The dynamics of a nuclear reaction is often considered in terms of the time evolution of selected degrees of freedom, such as distance between mass centers of nuclei, different kinds of deformation, mutual orientation, mass and charge asymmetry, etc.~\cite{zagrebaev2015}. This evolution is controlled by the nucleus-nucleus interaction potential which depends on all the chosen degrees of freedom.
    
    At the same time, the most important characteristic of the interaction potential is its dependence on the center-of-mass distance. It is characterized by a barrier (the Coulomb/fusion barrier) usually calculated in an assumption of spherical shapes of nuclei. The dependence of nuclear potentials on other degrees of freedom can be considered as a distortion of spherical nuclear shape. Thus, the calculation and discussion of the interaction potential for two spherical nuclei are of special importance. Considering the initial (approaching) stage of any nuclear reaction, it is necessary to know the position and the height of the Coulomb barrier. In some cases, the Coulomb barrier height can be derived from experimental data (e.g. fusion cross sections). In this regard, the semi-empirical Bass potential~\cite{Bass1977} is frequently used as “experimental” one since its parameters are fitted to reproduce available experimental data on the Coulomb barriers.
    Different parameterizations of the Coulomb barrier are studied in works~\cite{GHODSI2013,MANJUNATHA2018}; calculations of the barrier profiles obtained within various nuclear interaction models are presented in Refs.~\cite{Zagrebaev2007,qu2014,Back2014}.
    
    The description of the Coulomb barrier is important not only for near-barrier fusion reactions and multinucleon transfer reactions. Deep sub-barrier reactions are also a subject of theoretical studies since they play an important role in nucleosynthesis in the Universe~\cite{Zagrebaev2007_astro}.
    An estimate for the one-dimensional Coulomb barrier can be obtained within the framework of the folding procedure~\cite{Zagrebaev2007}, which requires calculating the nucleon density distribution in nuclei. The main geometrical parameters of the distributions --- the radius and thickness of the diffuse layer of nuclei --- determine the shape and position of the barrier. In this regard, the accuracy of these parameters requires special attention. This can be achieved based both on theoretical modeling and experimental data~\cite{Adamian2016,Sukhareva2021,Chamon2002}.
    
    The primary purpose of this work is the development of a nucleus-nucleus potential model compatible with available experimental data and suitable for its further application in nuclear reaction studies. We are interested here in the behavior of the nucleus-nucleus potential up to the moment of contact. The specifics of the potential in more compact configurations (for instance, repulsive core) or the adiabatic collision mode are beyond the scope of the present study.
    
    In this paper, we consider experimental data on the neutron and proton distribution in nuclei and propose the most suitable approximations for their radii and diffusenesses. We investigate how the nucleon density parameters influence the position of the Coulomb barrier derived from the folding potential. Finally, we employ Bass barriers as “experimental” ones and make some modifications to our model in order to reproduce these semi-empirical barriers.

\section{Form of the nucleon density distribution \label{sec:dens_params}}

    A range of analytical formulas is developed to describe the matter, pointlike nucleon, and charge distributions in atomic nuclei. These formulas are Fermi distribution, Gaussian model and sum of Gaussians, Fourier--Bessel expansion, and some other equations \cite{DeVries1987,Hasse1988}. The most exact fit of the densities extracted from experimental charge form factors is provided by expansions (sum of Gaussians, Fourier--Bessel) but these equations include a large number of free parameters. To study general aspects of nucleon distributions, one should consider equations with few free parameters. Since we are interested in heavy ions and the one-dimensional Coulomb barrier, here we take into account only spherical nuclei with $Z, N \geqslant 8$ to avoid the strong impact of shell and cluster effects on nucleon density characteristics.
    
    The simplest and the most universal form of nucleonic density is the Fermi representation. This distribution can be written as a standard 2-parameter equation or in the form taking into account possible central depression with 3 \cite{DeVries1987} or 7 \cite{Abdulghany2018} independent parameters. The symmetrized Fermi distribution with useful analytical features is also suggested \cite{Hasse1988}. We chose a basic 2-parameter form
    
    \begin{equation}\label{eq:fermi-distribution}
     \rho(r)=\cfrac{\rho_0}{1+\exp\left(\cfrac{r-R_0}{a}\right)},
    \end{equation}
    where $a$ is the diffuseness parameter, $R_0$ is the half-density radius, $r$ is the radius in spherical coordinates. The constant $\rho_0$ is determined by normalization:
    \begin{equation}\label{eq:normalization}
        X = \int \rho(r) \, d^3r,
    \end{equation}
    where $X$ can be equal to the nuclear charge~$Z$, number of neutrons~$N$, or mass number~$A$. Evidently, $\rho(0)$ differs from the constant $\rho_0$ only for the nuclei with $Z < 8$. For the nuclei under consideration, one may suppose that $\rho(0)\approx\rho_0$. For instance, in the case of $Z\sim10$, one obtains the approximation $\rho(0)\approx0.994\rho_0$ (when $R_0=2.8$~fm, $a=0.55$~fm), that confirms the assumption.

    The ``volume'' density (also called folded)~-- matter or charge~-- is expressed in terms of the pointlike density with a folding procedure:
    \begin{equation}\label{eq:density_folding}
        \rho_{volume}(r)=\int f(\mathbf{|r-r'|})\rho_{point}(r') \, d^3r',
    \end{equation}
    where $r'$ is the radial coordinate of the point nucleon, $f(\mathbf{|r-r'|})$ is the distribution of charge or matter in the nucleon. By the ``volume'' distribution, we mean the distribution of nucleons that takes into account nonzero sizes of nucleons. These are extracted from experimental measurements. Note that the matter distribution in the single neutron is considered equal to the charge distribution in the proton~\cite{Chamon2002}.
    
    Unobservable pointlike nucleon distributions are required to calculate the nucleus-nucleus interaction potential. We base our estimations on charge density data and thus extract pointlike distribution from the charge distribution. We achieve this using the relations between density parameters. The root-mean-squared (rms) radius is defined as usual:
    \begin{equation}\label{eq:rms_radius}
        \langle r^2 \rangle^{1/2}=\left(\frac{1}{X}\int r^2 \rho(r) \, d^3r\right)^{1/2},
    \end{equation}
    where $X = Z, N$ or $A$ and $\rho$ is the pointlike or matter density corresponding to the given rms radius. If density is described by the Fermi distribution~(\ref{eq:fermi-distribution}), then the half-density radius $R_0$ can be related to the rms radius on the assumption of $\exp{(\langle r^2 \rangle^{1/2}/a)}\gg 1$ which applies for all nuclei with $Z \geqslant 8$: 
    \begin{equation}\label{eq:r_0_approximation}
      R^2_0 \approx \langle r^2 \rangle  \left[ \frac53  - \frac73 \left( \frac{\pi a}{ \langle r^2 \rangle}\right)^2 \right].
    \end{equation}
    Consequently, the half-density radius equal to the rms one in the case
    \begin{equation}
       \langle r^2 \rangle =  \frac72 (\pi a)^2.
    \end{equation}
    Choosing the average diffuseness value of 0.55~fm, one obtains
    \begin{equation*}
       \langle r^2 \rangle^{1/2}= \sqrt{\frac72}  \pi a \approx 3.2~\text{fm},
    \end{equation*}
    that corresponds to stable nuclei with $Z=15,16$. 

    Based on the folding equation~(\ref{eq:density_folding}), one can express the relation between the rms radii of charge and nucleon distributions taking into account nucleons sizes~\cite{Friar1975}:
    \begin{equation}\label{eq:radii_ch_p}
        \langle r^2 \rangle_{ch}=\langle r^2 \rangle_{p}+\langle r^2 \rangle_{proton}+\frac{N}{Z}\langle r^2 \rangle_{neutron},
    \end{equation}
    where $\langle r^2 \rangle_{proton},\langle r^2 \rangle_{neutron}$ is the mean-squared charge radius of the proton or the neutron. The $\langle r^2 \rangle_{neutron}$ radius has the conventional value: $-0.1161$~fm${}^2$ \cite{Zyla2020_PDG}. As for the rms charge radius of the proton $\langle r^2 \rangle_{proton}^{1/2}$, there are two average values confirmed with experimental data: 0.8751(61)~fm (CODATA2014, \cite{CODATA2014}) and 0.8414(19)~fm  (CODATA2018, \cite{CODATA2018}). The most recent works do not resolve this inconsistency, but new data and refinements seem to confirm the smaller CODATA2018 value of 0.8414~fm \cite{Karr2020,Gao2022}. Additionally, spin-orbit and relativistic  Darwin-Foldy terms can be included in the Eq.~(\ref{eq:radii_ch_p}) \cite{Friar1975}. However, they can be neglected as insignificant values that do not substantially contribute to the calculations. We use the value $\langle r^2 \rangle_{p}^{1/2}=0.841$~fm which is currently recommended by CODATA and PDG \cite{Zyla2020_PDG}.
    
    The normalization integral~(\ref{eq:normalization}) equals to proton number $Z$ in both proton and charge distribution. Due to the rms proton radius being smaller than the rms charge radius, the diffuseness needs to increase when transitioning from the charge distribution to the distribution one assuming that all distributions have Fermi-form~(\ref{eq:fermi-distribution}). This relation can be described by the equation~\cite{Lima2004}
    \begin{equation}\label{eq:diff_transformation}
        a_p=a_{ch}-0.03.
    \end{equation}
     
\begin{figure}[htb]
         \center{\includegraphics[width=\linewidth]{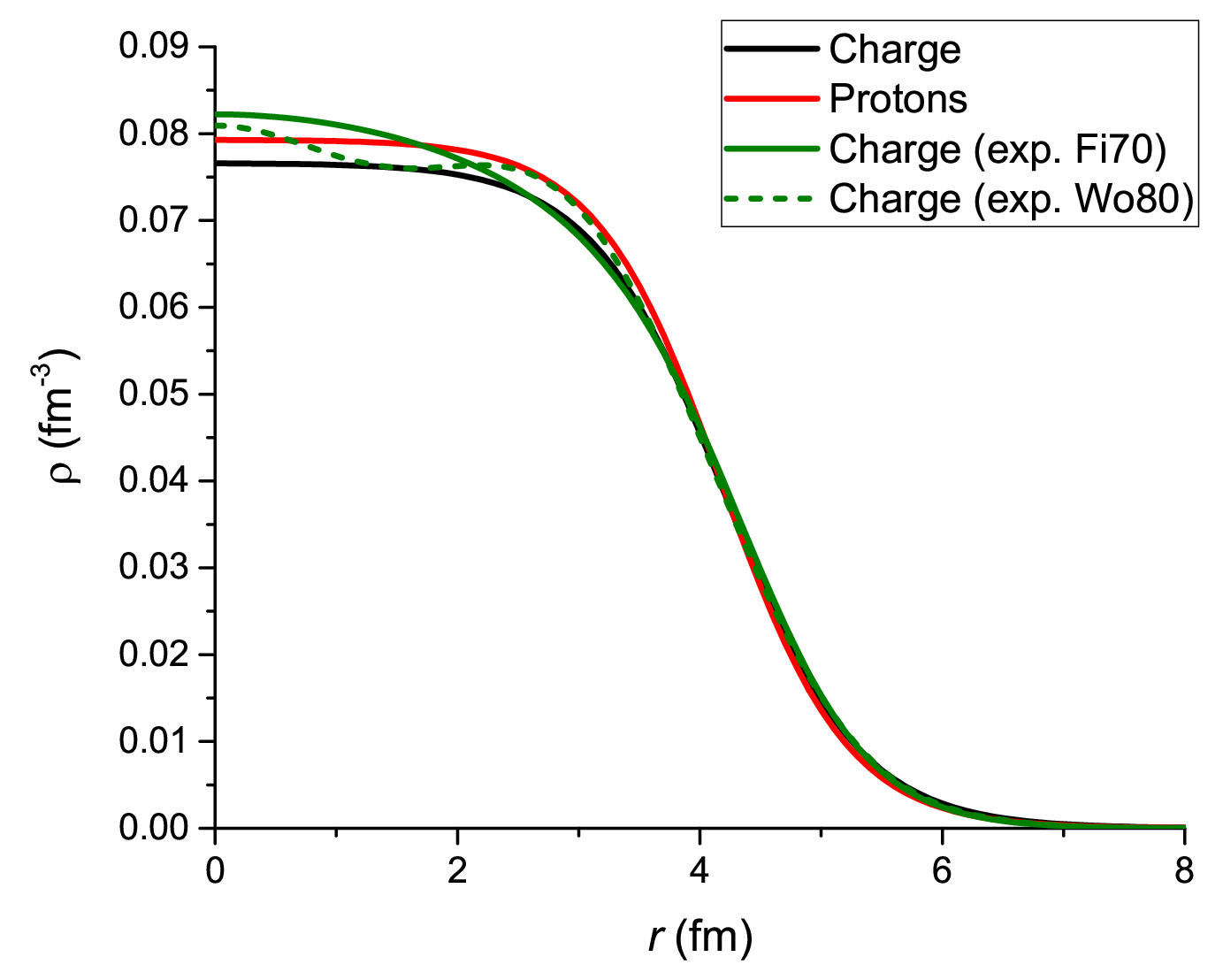}}
         \caption{\label{fig:Ni_prot_dens} Calculated proton (red) and charge (black) densities of ${}^{60}$Ni depending on the radial coordinate. The experimental charge density (green curves) is also shown for comparison: solid line (Fi70) corresponds to the 3-parameter Fermi form~\cite{Ficenec1970}, dashed line (Wo80)~--- Fourier-Bessel expansion~\cite{Wohlfahrt1980}.}

    \end{figure}

     An example of the pointlike proton density parameters determined using Eq.~(\ref{eq:r_0_approximation}), (\ref{eq:radii_ch_p}), (\ref{eq:diff_transformation}) is shown in Fig.~\ref{fig:Ni_prot_dens}. The charge density distribution in ${}^{60}$Ni has a slightly more diffuse edge than the proton density distribution and therefore has less central density. Half-density radii of proton and charge distributions have almost identical values because for ${}^{60}_{28}$Ni correction of $\langle r^2 \rangle_{proton}+\frac{32}{28}\langle r^2 \rangle_{neutron}\approx0.57$~fm${}^2$ is negligible in comparison with $\langle r^2 \rangle_{p}\approx\langle r^2 \rangle_{ch}\approx14.6$~fm${}^2$~\cite{Angeli2013}.
     
\section{Experimental data on nucleon density parameters \label{sec:exp_data}} 

    This section analyzes available experimental data on the nucleon density distribution. We rely mainly on the charge distribution data and employ theoretical models for additional verification since these data are much more extensive than those for the neutron density distribution.
    
    We investigated  characteristics of the charge distribution: the rms radius $\langle r^2 \rangle^{1/2}_{ch}$ and the diffuseness $a_{ch}$. The data on the neutron skin thickness $\mathrm{\Delta} r_{np}$ is also taken into account. This quantity is determined as the difference between  the rms radii of the proton and neutron pointlike distribution:
    \begin{equation}\label{eq:neutron_skin}
      \mathrm{\Delta} r_{np} = \langle r^2 \rangle_{n}^{1/2} - \langle r^2 \rangle_{p}^{1/2}. 
    \end{equation}
    The neutron skin thickness allow connecting data on the proton and the neutron distributions.
    
    \subsection{\label{subsec:charge_radii} Charge radii}
\begin{figure}
       \center{\includegraphics[width=\linewidth]{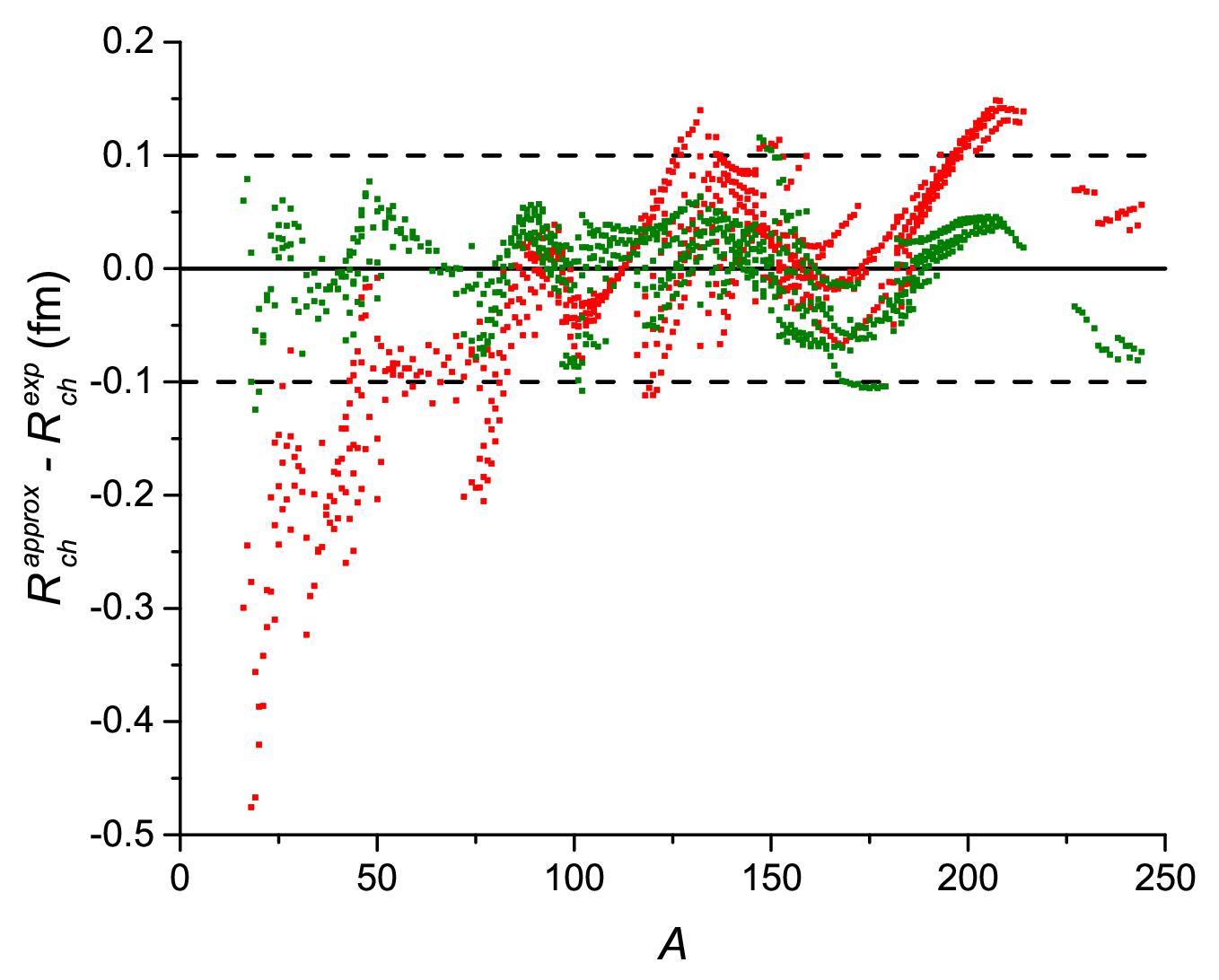}}
       \caption{\label{fig:radii_residuals} Difference between estimations and experimental values of the charge rms radii depending on the mass number $A$. Approximation data~(\ref{eq:r_NP}) are marked by green squares, values of standard fit  $R_{ch}=r_0 A^{1/3}$ are marked by red squares.}
    \end{figure}

    Our knowledge about nucleus charge distributions is based on electron scattering experiments and experiments with muonic atoms ~\cite{DeVries1987}.
    A limited set of experimental data on charge density parameters in the form of a 2-parameter Fermi distribution can be found in Ref.~\cite{DeVries1987}: the diffuseness and half-density radii are listed there for 66 nuclei with $9 \geqslant Z \geqslant 92$. In addition to the experiments mentioned above, X-ray and optical spectroscopy allow measuring the difference in the rms charge radii along isotopic chains~\cite{Angeli2013,Hammen2018,DeGroote2020}. These techniques significantly expand available data on charge radii. Excluding estimations for unstable isotopes, one can find the rms charge radii values for 836 isotopes in Ref.~\cite{Angeli2013}. To outline global trends of radii variation depending on $N,Z,A$, we considered analytical formulas available in the literature \cite{Bayram2013,Angeli2004}. All formulas are fitted on the data from Ref.~\cite{Angeli2013} (813 nuclei with $Z\geqslant8, A\geqslant16$). We found that the following approximation, initially suggested in Ref.~\cite{Nerlo-Pomorska1994}:
    \begin{equation}\label{eq:r_NP}
        R_{ch}=\langle r^2 \rangle_{ch}^{1/2}=r_0 \left(1-b I+c \frac{1}{A}\right) A^{1/3},
	\end{equation}
    where neutron excess is denoted as
    \begin{equation*}
        I=\frac{N-Z}{A},
	\end{equation*}
    indicates the smallest rms deviation of 0.41~fm.  Newly established coefficients are $r_0=0.9560(14)$~fm, $b=0.1527(67)$, $c=2.326(63)$. Figure~\ref{fig:radii_residuals} demonstrates that discrepancies between estimations with Eq.~(\ref{eq:r_NP}) and experimental data do not exceed 0.1~fm. The values obtained with the standard formula $R_{ch}=r_0 A^{1/3}$ with $r_0=0.952$~fm (fitted on data from Ref.~\cite{Angeli2013}) are also shown. The comparison demonstrates that additional neutron excess and mass corrections in Eq.~(\ref{eq:r_NP}) take into account macroscopic influences. Residual deviations from experimental data can be attributed to microscopic shell effects.  
    
    \subsection{\label{subsec:diffuseness} Diffuseness}
\begin{figure}
       \center{\includegraphics[width=\linewidth]{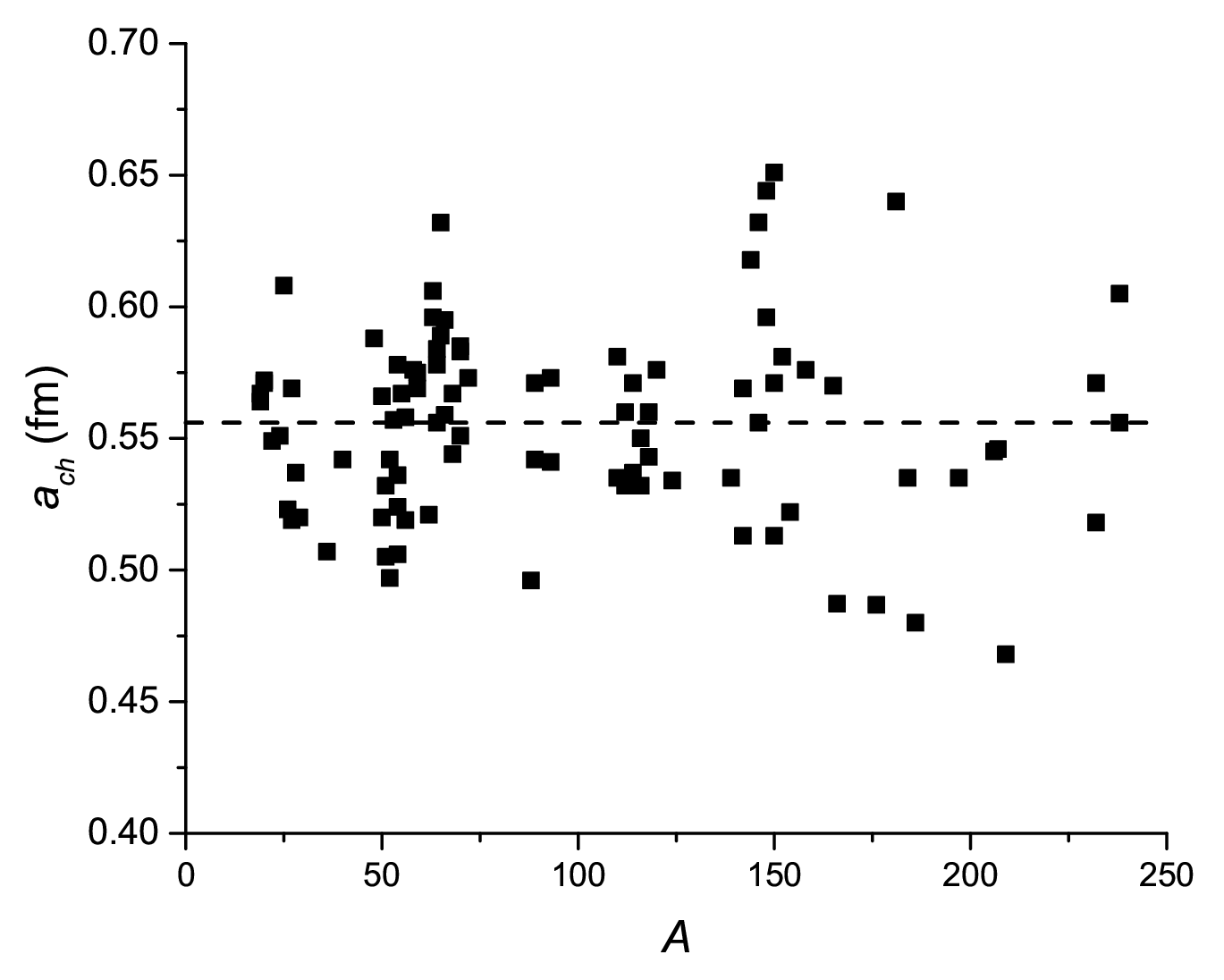}}
       \caption{\label{fig:diffuseness} Experimental diffuseness values~\cite{DeVries1987} depending on the mass number $A$. The average value of 0.556(4)~fm is shown by the dashed line.}
    \end{figure}

    The most complete set of experimental data on the diffuseness of the charge density is collected in Ref.~\cite{DeVries1987} as illustrated by Fig.~\ref{fig:diffuseness}. Some novel and reanalyzed experimental data can be found in Refs.~\cite{Abdulghany2018,Fricke1995}. There seems to be no clear correlation between diffuseness value and $N,Z,A$ numbers of corresponding nuclei. Besides, diffuseness estimations for one nucleus can differ from each other by $0.3-0.5$~fm. 
    Due to significant uncertainty of diffuseness, we use an approximate average value of 0.55~fm for the first stage calculation.
    
    The surface diffuseness of interacting nuclei significantly impacts on nuclear fusion because it determines the outcome result of the competition between the Coulomb and nuclear interactions. Therefore, the location of the  Coulomb barrier strongly correlates with the diffusenesses of colliding nuclei. We will discuss an effective way to adjust diffuseness in heavy-ion reactions in Sect.~\ref{subsec:diff_from_bass}.

    \subsection{\label{subsec:neutron_skin} Neutron skin thickness}
     Compared to the quantities of the charge distribution, the thickness of a neutron skin is more complicated to measure, first of all, due to the complexity of nuclear interaction. Experimentally, the neutron skin thickness is measured by extracting the information from the studies on electromagnetic phenomena: parity violation~\cite{Horowitz2012} and excitation of pygmy, giant, and spin-dipole resonances~\cite{Krasznahorkay1999,Klimkiewicz2007,Rossi2013} are investigated. Another type of experiments exploits the nuclei-antiproton scattering and combines it with spectroscopy of the annihilation residues and antiprotonic X-ray radiation~\cite{Trzcinska2001a,JASTRZEBSKI2004}. 
     
\begin{figure}[htb]
        \center{\includegraphics[width=\linewidth]{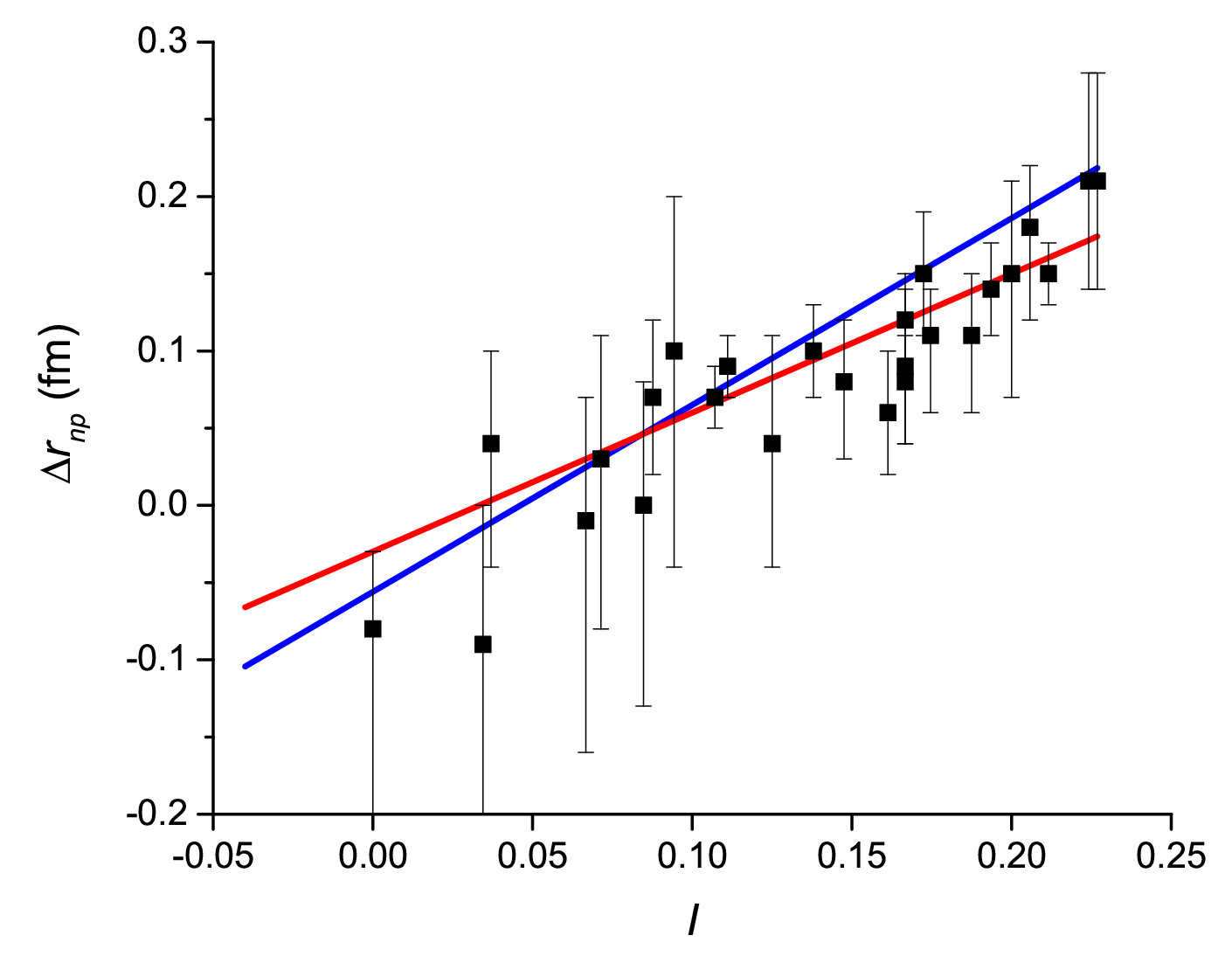}}
        \caption{\label{fig:neutron_skin} The neutron skin thickness dependence on the neutron excess $I$: experimental data (points) and linear fit (red line) from Ref.~\cite{JASTRZEBSKI2004}. Blue line corresponds to Skyrme--Hartree--Fock calculations with SLy4 forces.}
     \end{figure}

     It should be noted that the measurement of the neutron skin thickness implies substantial uncertainties (see Fig.~\ref{fig:neutron_skin})       
     and some discrepancies among the values obtained in different studies~\cite{Zhang2021}. For example, the neutron skin thickness of  ${}^{208}$Pb, a nucleus that is important for understanding nuclear matter structure, is estimated from 0.08(5) to 0.42(20)~fm with an average of 0.2~fm \cite{Zhang2021}. Nevertheless, the data extracted uniformly from an experiment reveals the trend of the neutron skin thickness variation from one nucleus to another. In Ref.~\cite{JASTRZEBSKI2004}, the neutron skin thickness is deduced from the antiprotonic atom X-ray data and described as the following linear dependency on the neutron excess $I$:
     \begin{equation}\label{eq:neutr_skin_approx}
        \mathrm{\Delta} r_{np}= a  I + b, 
     \end{equation}
     where $a=0.90(15)$~fm, $b=-0.03(2)$~fm. The graph for  Eq.~(\ref{eq:neutr_skin_approx}) is shown in Fig.~\ref{fig:neutron_skin}.  If $I=0$, the neutron skin thickness is negative according to approximation~(\ref{eq:neutr_skin_approx}). This can be explained by the Coulomb repulsion between protons and is confirmed by some experimental data for nuclei with $I\approx0$ (for example, see data for Ca and Ni isotopes in Ref.~\cite{Zhang2021}).

     We also calculated $\mathrm{\Delta} r_{np}$ following the Skyrme--Hartree--Fock approach with a SLy4 parameter set~\cite{Chabanat1998}. The neutron skin thickness was estimated for even-even isotopes in chains of O, Ca, Ni, Zr, Sn, Sm, Pb and fitted with linear dependency. We obtained $a = 1.21(3)$~fm, $b = - 0.058(6)$~fm in Eq.~(\ref{eq:neutr_skin_approx}).
     As Fig.~\ref{fig:neutron_skin} illustrates, lines corresponding to our coefficients and the coefficients in Eq.~(\ref{eq:neutr_skin_approx}) are close to each other. Microscopic calculations demonstrate a similar correlation between the neutron skin thickness and the neutron excess~\cite{Bespalova2017,Zhang2021}.

\subsection{\label{sec:p_n_densities} Proton and neutron densities}
    
        First, we tested what density distributions can be obtained by applying the equations described above: for the half-density radius~(\ref{eq:r_0_approximation}), for the neutron skin thickness~(\ref{eq:neutron_skin}), and for the rms charge radius~(\ref{eq:r_NP}). Let us give some examples of the calculated density distributions. We fixed the value of charge distribution diffuseness as $a_{ch}=0.55$~fm. Consequently, the diffuseness value for proton distribution $a_p=0.52$~fm from Eq.~(\ref{eq:diff_approx}) was obtained. We assumed that diffusenesses of the proton and neutron distributions were equal since nuclei under consideration are supposed to have no halo-type structure of nuclear distributions~\cite{Thiel2019}. Half-density radii $R_p, R_n$ were determined from~(\ref{eq:r_0_approximation}), (\ref{eq:neutron_skin}),  (\ref{eq:r_NP}) equations.
         
        We present the results of calculations based on approximated data for density parameters in  Tables~\ref{tab:rho_diff}, \ref{tab:radii}.
    \begin{table}[bt]
        \caption{Density parameters for spherical nuclei: normalization constant $\rho_0$ (fm$^{-3}$) and diffuseness $a$ (fm). In the upper part of the table, 8 spherical nuclei used to calculate the approximation~(\ref{eq:diff_approx}) are listed; data for ${}^{86}$Kr and ${}^{136}$Xe not used for fitting are also included. Diffusenesses $a_p$ correspond to Eq.~(\ref{eq:diff_approx}). Experimental diffusenesses of the charge distribution are given for reference.     \label{tab:rho_diff} }
        \begin{ruledtabular}
        \begin{tabular}{ccccccc}
            Z  & A   & $\rho_{0p}$ & $\rho_{0n}$ & $\rho_{0tot}$ & $a_p$ & $a_{ch}$ (exp)   \\    
        \hline
            8  & 16  & 0.100  & 0.106  & 0.206   & 0.56\footnotemark[1]        &       \\
            20 & 40  & 0.087  & 0.090  & 0.177   & 0.59\footnotemark[1]           & 0.613~\cite{Abdulghany2018} \\
            20 & 48  & 0.078  & 0.096  & 0.173   & 0.59\footnotemark[1]          &       \\
            28 & 60  & 0.079  & 0.088  & 0.167   & 0.60\footnotemark[1]           & 0.548~\cite{Abdulghany2018} \\
            40 & 90  & 0.073  & 0.087  & 0.160   & 0.62\footnotemark[1]         &       \\
            50 & 124 & 0.067  & 0.090  & 0.157   & 0.63\footnotemark[1]          & 0.534~\cite{DeVries1987} \\
            62 & 144 & 0.069  & 0.085  & 0.154   & 0.65\footnotemark[1]          &       \\
            82 & 208 & 0.063  & 0.088  & 0.152   & 0.66\footnotemark[1]        & 0.544~\cite{Abdulghany2018} \\
            \hline
            36 & 86  & 0.072  & 0.090  & 0.162   & 0.60\footnotemark[2]           &       \\
            54 & 136 & 0.066  & 0.090  & 0.156   & 0.60\footnotemark[2]          &      \\
        \end{tabular}
        \end{ruledtabular}
        \footnotetext[1]{Diffuseness value in pair with ${}^{208}$Pb.}
        \footnotetext[2]{Diffuseness value in pair of ${}^{86}$Kr and ${}^{136}$Xe.}
        \end{table}
        Table~\ref{tab:rho_diff} shows central density constants $\rho_{0p}$, $\rho_{0n}$ for proton  and neutron distributions. In Table~\ref{tab:radii}, the half-density and rms radii are listed (denoted as ``Old''). It should be noted that the rms charge radius of both ${}^{48}$Ca and ${}^{16}$O deviates from the experimental value by about 2\%. The accuracy of charge radii prediction is higher for other spherical nuclei listed in Tab.~\ref{tab:radii} because density depression caused by shell effects is less pronounced. 
         
\begin{figure}[h!bt]
            \center{\includegraphics[width=\linewidth]{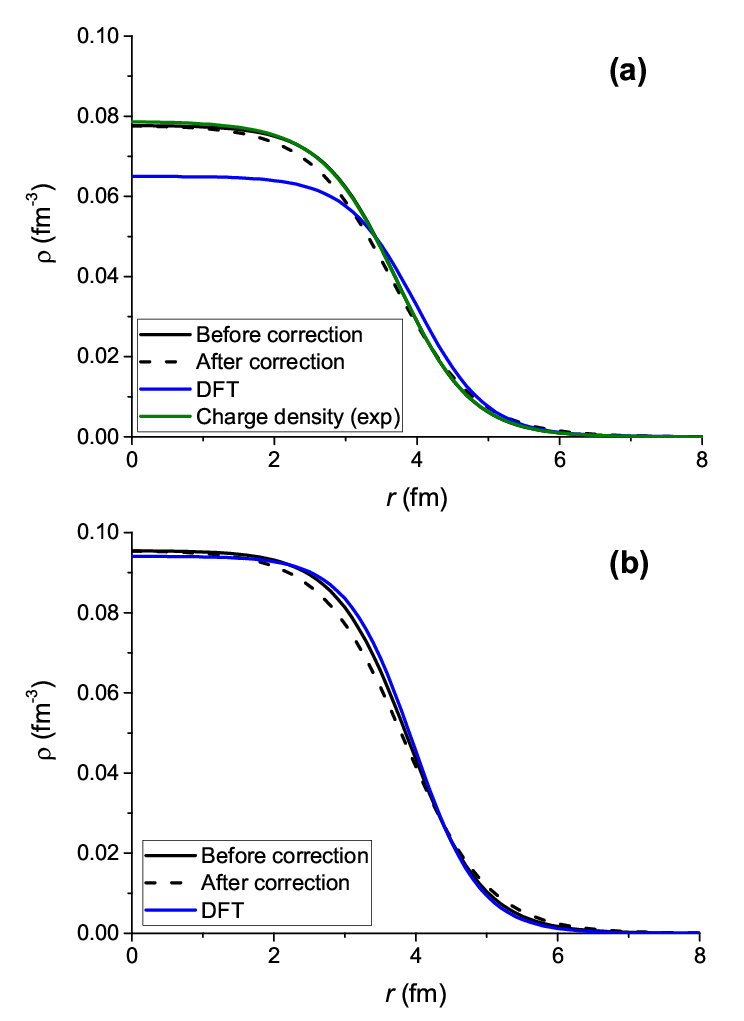}}
             \caption{\label{fig:Ca_prot_dens} Radial density distribution in ${}^{48}$Ca: a) proton and charge densities, b) neutron density. Results of the calculation before and after diffuseness correction are shown with solid and dashed black lines, respectively. Theoretical proton density (DFT)~\cite{Adamian2016} is denoted by the blue line, experimental charge distribution~\cite{Bellicard67} --- by the green line.}        
        \end{figure}

        Proton and neutron density distributions for ${}^{48}$Ca (denoted as ``Before correction'') are shown in Fig.~\ref{fig:Ca_prot_dens}.
        The experimental charge density is very close to the proton density calculated with our equations. The proton density predicted by the density functional theory (DFT) in Ref.~\cite{Adamian2016} has insufficient saturation in the central region of the nucleus. As for neutron densities, our calculation mostly corresponds to those in Ref.~\cite{Adamian2016}. 

        \begin{table*}[hbt]
        \caption{\label{tab:radii} Radial quantities (in fm): nucleon half-density $R_{0p}$, $R_{0n}$, and rms $\langle r^2\rangle_p^{1/2}$, $\langle r^2\rangle_n^{1/2}$ radii. Estimations of the charge rms radii $\langle r^2\rangle_{ch}^{1/2}$ are obtained with Eq.~(\ref{eq:r_NP}),   experimental values are taken from Ref.~\cite{Angeli2013}. ``New'' radii correspond to the diffusenesses from Table~\ref{tab:rho_diff}, ``old'' radii correspond to $a_p=0.52$~fm.}
        \begin{ruledtabular}
        \begin{tabular*}{\linewidth}{cccccccccccl}
        \multirow{2}{*}{Z}  & \multirow{2}{*}{A}  & \multicolumn{2}{c}{$R_{0p}$}& \multicolumn{2}{c}{$R_{0n}$} & \multicolumn{2}{c}{$\langle r^2\rangle_p^{1/2}$} & \multicolumn{2}{c}{$\langle r^2\rangle_n^{1/2}$} & \multicolumn{2}{c}{$\langle r^2\rangle_{ch}^{1/2}$} \\
         \cline{3-4}  \cline{5-6}  \cline{7-8}  \cline{9-10}  \cline{11-12} 
         & & Old  & New  & Old  & New  & Old  & New  & Old  & New  &   This work   &   \text{Exp}  \\
        \hline
        8  & 16  & 2.34 & 2.29 & 2.28 & 2.23 & 2.65 & 2.74 & 2.62 & 2.71 & 2.76 & 2.699(5)   \\
        20 & 40  & 3.57 & 3.50 & 3.52 & 3.46 & 3.37 & 3.49 & 3.34 & 3.46 & 3.46 & 3.4776(19) \\
        20 & 48  & 3.72 & 3.66 & 3.91 & 3.85 & 3.47 & 3.58 & 3.59 & 3.69 & 3.55 & 3.477(2)   \\
        28 & 60  & 4.18 & 4.11 & 4.23 & 4.16 & 3.77 & 3.89 & 3.80 & 3.92 & 3.85 & 3.8225(19) \\
        40 & 90  & 4.89 & 4.82 & 4.99 & 4.92 & 4.25 & 4.39 & 4.32 & 4.45 & 4.32 & 4.269(1)   \\
        50 & 124 & 5.46 & 5.39 & 5.67 & 5.60 & 4.65 & 4.79 & 4.80 & 4.93 & 4.72 & 4.674(2)   \\
        62 & 144 & 5.85 & 5.77 & 5.98 & 5.90 & 4.93 & 5.07 & 5.02 & 5.16 & 4.99 & 4.952(3)   \\
        82 & 208 & 6.63 & 6.56 & 6.86 & 6.78 & 5.49 & 5.64 & 5.65 & 5.79 & 5.54 & 5.5012(13) \\
        \hline
        36 & 86 & 4.75 & 4.69 & 4.92 & 4.86 & 4.16 & 4.27 & 4.27 & 4.38 & 4.23 & 4.184(2)   \\
        54 & 136 & 5.65 & 5.60 & 5.87 & 5.82 & 4.78 & 4.88 & 4.94 & 5.03 & 4.85 & 4.796(5)  \\
        \end{tabular*}
        \end{ruledtabular}
        \end{table*}

        Thus, the calculated densities demonstrate a close correspondence with experimental data in the case of proton density and with theoretical models in the case of neutron density. Next, we considered how these densities can be used to estimate the nucleus-nucleus interaction and evaluate the position of the Coulomb barrier.
    
\section{Double-folding potential \label{sec:diabatic_potential}} 

The evolution of a system of approaching nuclei is largely determined by the collision energy. Usually, two extreme modes of low-energy nuclear reactions are considered depending on the velocity of the projectile nucleus: adiabatic and diabatic. For near-barrier collisions, nuclei enter the range of nuclear forces quite slowly compared to the Fermi velocity. They almost stop at the contact distance and only intranuclear motion affects further process of nucleon exchange between nuclei. Such a collision mode is called adiabatic. The adiabatic potential is difficult to calculate because necessitates taking into account the internal structure of nuclei, which is dynamically transformed.
    
Collision in the diabatic mode suggests that the interaction of nuclei occurs rather quickly for the distribution of nucleons to reach equilibrium. In such reactions, the densities can be considered unchanged. Diabatic potential with the ``frozen'' nucleon distributions in nuclei is applicable to the above-  and sub-barrier reactions with heavy ions. However, if the distance between approaching nuclei is more than the sum of their half-density radii, nucleon density in the overlap area is less than the average central density $\rho_{0tot}$ of the reacting nucleus. Under this condition, the repulsion caused by the Pauli exclusion principle for nucleons is relatively weak so that diabatic and adiabatic potentials are identical. This fact explains why diabatic potential can be used to estimate the Coulomb barrier position. It should be noted that if one uses the diabatic potential to describe near-barrier reactions, then a correction of the density parameters may be required since such a process is not strictly diabatic.
    
\begin{figure}
    \center{\includegraphics[width=\linewidth]{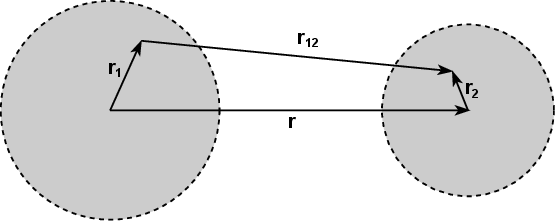}}
    \caption{\label{fig:radii_diag} A schematic picture of two spherical nuclei.}
\end{figure}

The most consistent way to calculate diabatic potential $V$ is to use the folding procedure when nucleon-nucleon interaction potential $v$ is averaged over nucleon density distributions. In the case of central collision of two spherical nuclei, it is expressed as 
\begin{equation}\label{eq:folding}
    V(r) = \int \rho_1(r_1) \int \rho_2 (r_2) v(r_{12}) \, d^3 r_1 \, d^3 r_2,
\end{equation}    
where $r_{12}=|\mathbf{r+r_2-r_1}|$ is the distance between nucleons, $r_1, r_2$ are the radii in a coordinate system of corresponding nuclei, $r$ is the distance between centers of nuclei, $\rho_{1,2}$ are the total nucleon densities. Figure~\ref{fig:radii_diag} clarifies the relative position of the vectors $\mathbf{r,r_1,r_2,r_{12}}$. Total nucleon densities include proton and neutron pointlike densities:   
\begin{equation}\label{eq:sum_of_densities}
     \rho_{tot}(r)=\rho_p(r)+\rho_n(r).
\end{equation}
	
The internucleon interaction potential $v$ consists of two parts, the Coulomb and nuclear interaction:
\begin{equation}\label{eq:nucl_coul}
    v(r)=v_{nucl}(r)+v_{Coul}(r).
\end{equation}
where Coulomb term is written as
\begin{equation}\label{eq:pot_coul}
    v_{Coul}(r)=\frac{e^2}{r}.
\end{equation}
To describe internucleon interaction, we use the Migdal potential~\cite{Migdal1983}
\begin{multline}\label{eq:pot_Migdal}
    v_{nucl}(\mathbf{r_1, r_2}) \\
    = C\left[ F_{ex}+(F_{in}-F_{ex})\frac{\rho_1(r_1)+\rho_2(r_2)}{\rho_{00}}\right]\delta(\mathbf{r_{12}}),
\end{multline}
where
\begin{equation}\label{eq:f_params}
    F_{ex,in}=f_{ex,in}\pm f'_{ex,in}.
\end{equation}
The sign ``$+$'' corresponds to the interaction of identical nucleons ($pp$ and $nn$ interaction), ``$-$'' corresponds to the proton-neutron interaction; indices \textit{ex,in} refer to the ``interior'' region of the nuclear matter distribution, where nucleon density is high, and to the ``exterior'' region where density is close to zero. Parameter $\rho_{00}$ in Eq.~(\ref{eq:pot_Migdal}) equals the average central density of two interacting nuclei: $\rho_{00}=0.5(\rho_{01}+\rho_{02})$, where $\rho_{0i}$ ($i=1,2$) is the normalization parameter for total nucleon density that is equal to the sum of proton and neutron densities. Other parameter values are the following~\cite{Migdal1983}: $f_{in} = 0.09, f_{ex} = -2.59, f'_{in}= 0.42$, and $f'_{ex} = 0.54$. 

The normalization constant $C$ in Eq.~(\ref{eq:pot_Migdal}) depends on the density of states at the Fermi surface~\cite{Speth2014}. Without going into details, let us note that $C$ is connected to the Fermi momentum $p_{F}$ and, consequently, to the saturated nucleon density $\rho_{00}$~\cite{Migdal1983,Speth2014}:
	$$C \sim \frac{1}{p_F} \sim \frac{1}{\sqrt[3]{\rho_{00}}} .$$
Based on the Migdal reference value $C^*=300$~MeV$\cdot$fm${}^3$ from Ref.~\cite{Migdal1983} which corresponds to $\rho^*_{00} = 0.17$~fm${}^{-3}$, we employ the following equation to calculate the factor $C$:
\begin{equation}\label{eq:C_norm}
    C = C^* \left(\frac{\rho^*_{00}}{\rho_{00}}\right)^{1/3}.
\end{equation}
According to our calculations (see Table.~\ref{tab:rho_diff}), the saturation density $\rho_{00}$ varies from 0.152 to 0.206~fm${}^{-3}$, which leads to the factor $C$ changing from 281~MeV$\cdot$fm${}^3$ (${}^{16}$O$+{}^{16}$O reaction) to 312~MeV$\cdot$fm${}^3$ (${}^{208}$Pb$+{}^{208}$Pb reaction). The difference from the reference value $C^*$ plays a significant role in the reactions between the lightest nuclei only: ${}^{16}$O, ${}^{40}$Ca, ${}^{48}$Ca, and ${}^{60}$Ni.
For example, in the ${}^{16}$O$+{}^{16}$O  reaction, changing in $C$ to 281~MeV$\cdot$fm${}^3$ leads to the potential minimum rising by 2.3~MeV with a total depth of 25~MeV, whereas the barrier is placed at the level of 10~MeV. In the case of the reaction ${}^{48}$Ca$+{}^{208}$Pb ($C - C^*$ = 4.6~MeV$\cdot$fm${}^3$), the position of the minimum changes by 0.5~MeV, which is negligible against a background of the Coulomb barrier of 177 MeV. Thus, when one considers reactions with heavier nuclei, the difference between $C$ and $C^*$ has almost no effect on the result of the potential calculation hence the usage of the reference value $C^*$ is sufficient.
Moreover, changing the parameterizations of the nucleon distribution diffuseness Eq.~(\ref{eq:diff_approx}) (see Sec.~\ref{subsec:diff_from_bass}) results in the elimination of the difference between the two variants of calculations.

The Bass potential $V_{Bass}$ is widely used to estimate the Coulomb barriers. Expression for this potential is the following~\cite{Bass1977}:
\begin{equation}\label{eq:pot_Bass}
    V_{Bass}(r)=\frac{Z_1 Z_2 e^2}{r}-\frac{R_1 R_2}{R_1 + R_2}g(\xi),
\end{equation}
where $\xi=r-(R_1+R_2)$ is the distance between surfaces of interacting nuclei. Nuclear half-density radii are estimated as $R_i=1.16 A_i^{1/3}-1.39A_i^{-1/3}$. The function $$g(\xi)=\left(A \exp{\frac{\xi}{d_1}}+B\exp{\frac{\xi}{d_2}}\right)^{-1}$$ has the following parameters: $A=0.03$~MeV${}^{-1}\cdot$~fm, \linebreak $B=0.0061$~MeV${}^{-1}\cdot$~fm, $d_1=3.3$~fm, and $d_1=0.65$~fm.

    \subsection{\label{subsec:const_diff} The case of constant diffuseness}    
\begin{figure}[h]
        \center{\includegraphics[width=\linewidth]{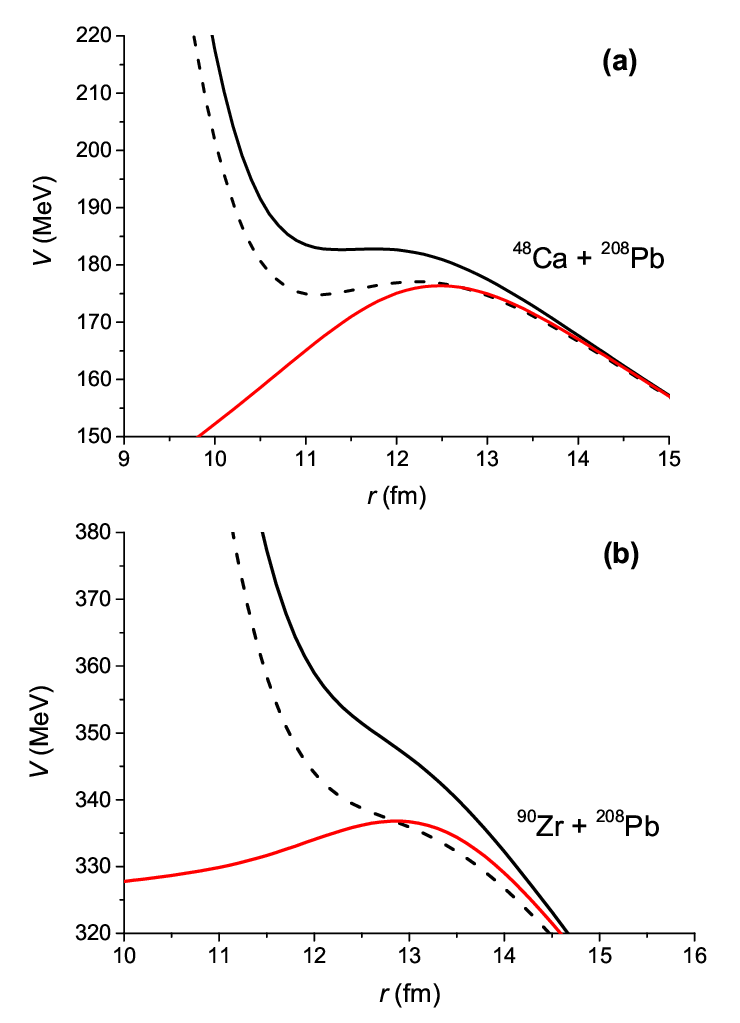}}
        \caption{\label{fig:Pb_Ca_potentials} Double-folding potentials depending on the distance between nuclei for the systems of a) ${}^{48}$Ca+${}^{208}$Pb and b) ${}^{90}$Zr+${}^{208}$Pb. Black lines correspond the double-folding potential: solid ``Before correction'' and dashed ``After correction'' lines. The Bass potential is depicted in red.}
    \end{figure}

   Nucleon density distributions described in the previous section represent distributions in noninteracting nuclei. If one adheres to the diabatic mode of collision, the folding potential should be derived from the distribution, diffuseness of which does not depend on the distance between nuclei. The numerical results of this approach for the systems of ${}^{48}$Ca+${}^{208}$Pb and ${}^{90}$Zr+${}^{208}$Pb are shown in Fig.~\ref{fig:Pb_Ca_potentials}.
   These potentials (denoted as ``Before correction'') are obtained using the diffuseness $a_p=a_n=0.52$~fm and demonstrate strong repulsion. A large difference between the calculated potential and Bass potential at the barrier (distance is about 12.5~fm for ${}^{48}$Ca+${}^{208}$Pb and 13~fm for ${}^{90}$Zr+${}^{208}$Pb) indicates that the chosen diffuseness value is too small so that the nuclear interaction is suppressed by the Coulomb force. Since the diffuseness varies from 0.45 to 0.65 even for noninteracting nuclei, we aimed to refine its values to reach a closer agreement between our results and the semi-empirical Bass barriers.

    \subsection{\label{subsec:diff_from_bass} Bass barrier as a reference point}    
    Although we make calculations with ``frozen'' densities, it is clear that more correct results for the Coulomb barrier can be obtained if the change in diffuseness is taken into consideration. 
    Moreover, diffuseness values of colliding nuclei dramatically influence the position and height of the barrier conditioned by an interplay between the Coulomb repulsion and nuclear attraction as it is shown in our previous work~\cite{Simonov2022}.
    
    In order to account for the time-independent distributions and consider the change in the nucleon distribution near the nuclei surfaces, an effective calculation method can be used. We assume that the central density does not change during the reaction. The constants $\rho_{0p},\rho_{0n}$ are taken from the calculation of the first stage (see Table.~\ref{tab:rho_diff}). Then we fit the barrier heights to reach a higher correspondence with the Bass barrier. Assuming that diffusenesses of two nuclei are equal and depend on a given pair of colliding nuclei, we adjust the heights and positions of the potential barrier by varying the values of diffuseness. 
     
\begin{figure}
        \center{\includegraphics[width=\linewidth]{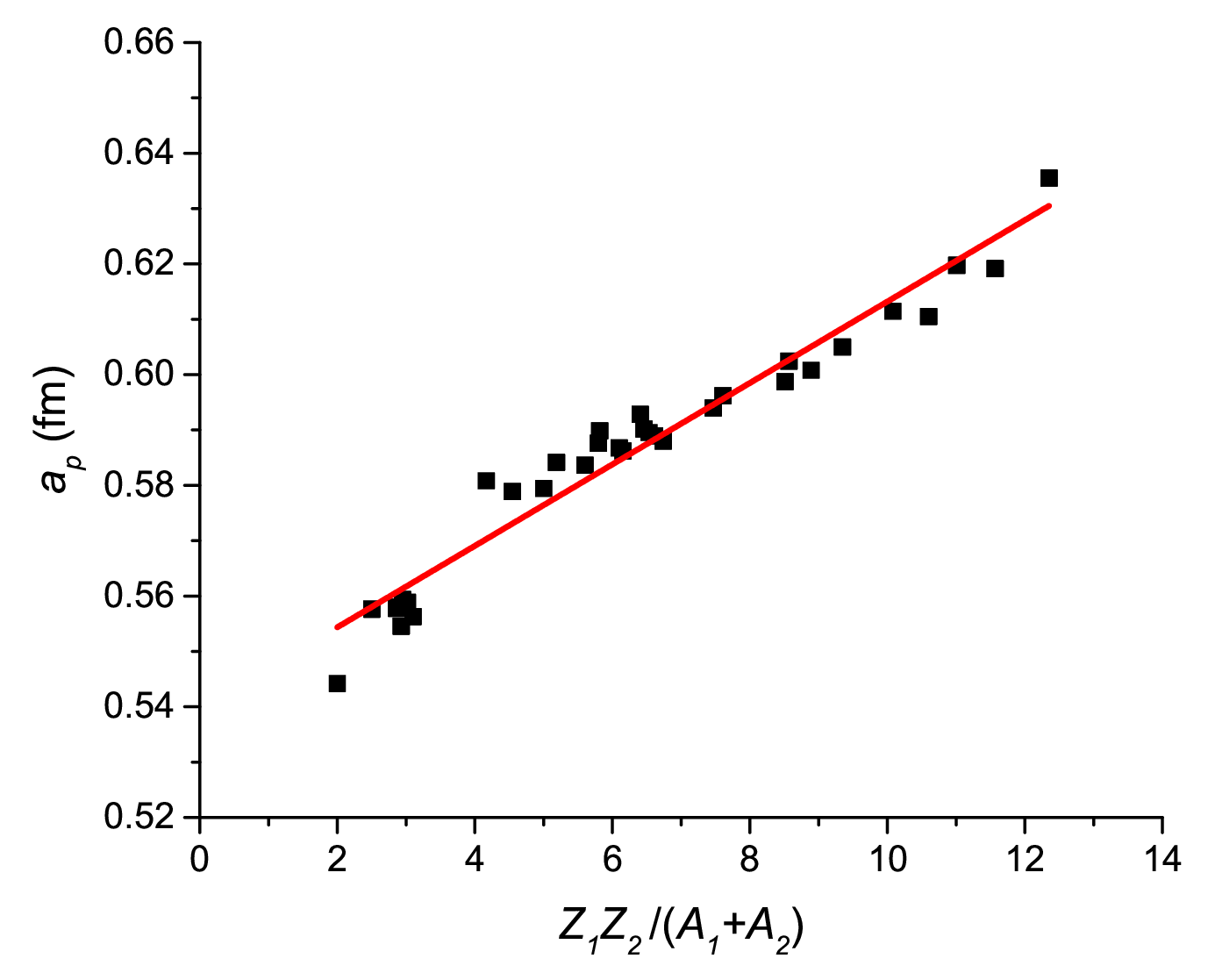}}
        \caption{\label{fig:diff_rho_const} New diffuseness values (points) obtained in the condition of the fixed central density $\rho_0$. The linear fit is shown by the red line.}
    \end{figure}

    At the second stage of calculations, we consider all nuclei pairs from the set of spherical nuclei with $Z\geqslant8$: ${}^{16}$O, ${}^{40}$Ca, ${}^{48}$Ca, ${}^{60}$Ni, ${}^{90}$Zr, ${}^{124}$Sn, ${}^{144}$Sm,  ${}^{208}$Pb. Since there is no potential maximum due to the very strong Coulomb repulsion for the three heaviest systems ${}^{144}$Sm~+ ${}^{144}$Sm, ${}^{144}$Sm~+ ${}^{208}$Pb, ${}^{208}$Pb~+ ${}^{208}$Pb, we obtain $(8\times8-2\times2)/2=30$ different pairs and diffuseness values. These values belong to the $0.53\div 0.65$~fm range (see Fig.~\ref{fig:diff_rho_const}).
    The obtained data can be described with the equation
	\begin{equation}\label{eq:diff_approx}
        a=a_0+a_1\frac{Z_1 Z_2}{A_1+A_2},
	\end{equation}
    where $a_0=0.5396(14)$~fm and $a_1=0.0074(2)$~fm; the Pearson correlation coefficient is 0.978, and the rms deviation is $4.4\cdot 10^{-3}$~fm. It is worth noting that for all nuclei new diffuseness values exceed the $0.52$~fm value which we have used at the first stage of our calculations (for new $a_p$ values  see Table~\ref{tab:rho_diff}). The form of the fraction $\frac{Z_1 Z_2}{A_1+A_2}$ indicates to the interplay between Coulomb and nuclear forces: the higher the $Z_1 Z_2$ product is, the more intense the Coulomb repulsion is, and the higher the potential barrier becomes. The sum of mass numbers in the term of the fraction is related to the number of nucleons and, consequently, to volume (and also surface) nuclear attraction. Therefore, it is necessary to achieve a balance between $Z_1 Z_2$ and $A_1+A_2$ factors to soften the potential barrier if  heavier isotopes react.
    
    The diffuseness has increased, but the half-density radius has decreased due to the normalization constant $\rho_0$ kept in the Eq.~(\ref{eq:normalization}). Since Bass barriers are lower than the diabatic barriers described in the previous subsection, new parameters of nucleon distribution must lead to a fuzzier edge of the reacting nuclei. As can be seen from Fig.~\ref{fig:Pb_Ca_potentials} (``After correction'' lines), the profile of interaction potential lowers as the role of nuclear forces increases. A weakly pronounced minimum appears in the case of the ${}^{48}$Ca+${}^{208}$Pb system.
    
\begin{figure}[hbt]
        \center{\includegraphics[width=\linewidth]{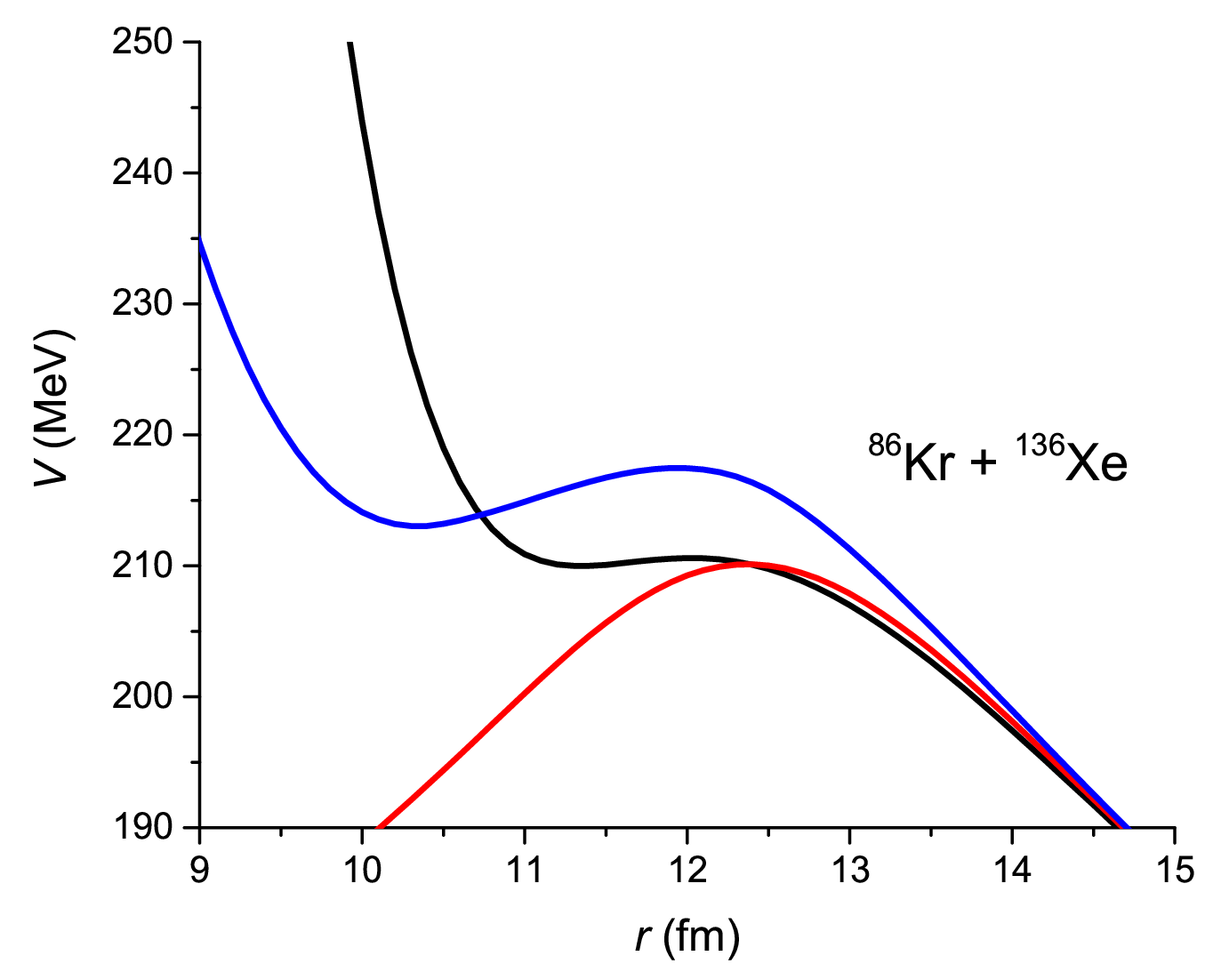}}
        \caption{\label{fig:pot_Kr_Xe} Nucleus-nucleus interaction potential for the system of ${}^{86}$Kr$+{}^{136}$Xe depending on the distance between nuclei. The double-folding potential is shown by the black line, the Bass potential~--- by the red line, and the proximity potential~--- by the blue line.}
    \end{figure}

    To demonstrate the capabilities of our approach, we consider a system of semi-magic nuclei \linebreak ${}^{86}$Kr~$+{}^{136}$Xe and calculate its interaction potential (Fig.~\ref{fig:pot_Kr_Xe}).
    The height of the potential barrier is close to the Bass barrier (the difference is less than 1~MeV), positions of the barriers differ by 0.5~fm. The proximity potential~\cite{Blocki1977} calculated with our radii values has a noticeable potential minimum which corresponds to a possible weakly bound state of the fusing system. However, the proximity potential barrier is too high to be used as an estimation of the barrier position.
    
    According to the analysis performed, we conclude that the approach described in this work can be implemented as a way to obtain estimates of the Coulomb barrier position in heavy-ion reactions. Experimentally-based approximations and relations between nucleon density parameters can be applied to describe density profiles in general and calculate interaction potential in low-energy heavy-ion reactions.

\section{Conclusion} 
The calculation of the nucleus-nucleus interaction potential in low-energy nuclear reactions is a relevant objective of theoretical physics. The double-folding method has proven to be a consistent method for calculating the nucleus-nucleus potential based on the distribution of nucleon densities. In this work, special attention was paid to the nucleon density distributions. Parameters of the neutron and proton densities were extracted from experimental data on the root-mean-square charge radius, the diffuseness of charge density distributions, and the neutron skin thickness. A new parameter set for the charge radii approximation was suggested. 

As a result of the data treatment, the diabatic potentials for various systems of heavy ions were calculated. It was shown that it is necessary to consider the change in the diffuseness value when reacting nuclei approach. The Bass barrier can be used as a reference point to take into account such change in the diffuseness of the nucleon distribution. The parameters of nucleon densities were refined using the relation obtained for diffusenesses of interacting nuclei. The exact theoretical prediction of the barrier height should be carried out using microscopic approaches, but the near-barrier region of nucleus-nucleus interaction potential dependence on the distance between nuclei and the position of the Coulomb barrier can be estimated with the extended approach based on experimental data.

\begin{acknowledgements}
Work of M. Simonov is supported by the grant of the Theoretical Physics and Mathematics Advancement Foundation ``BASIS''.
\end{acknowledgements}

\providecommand{\noopsort}[1]{}\providecommand{\singleletter}[1]{#1}%


\begin{thebibliography}{57}%
\makeatletter
\providecommand \@ifxundefined [1]{%
 \@ifx{#1\undefined}
}%
\providecommand \@ifnum [1]{%
 \ifnum #1\expandafter \@firstoftwo
 \else \expandafter \@secondoftwo
 \fi
}%
\providecommand \@ifx [1]{%
 \ifx #1\expandafter \@firstoftwo
 \else \expandafter \@secondoftwo
 \fi
}%
\providecommand \natexlab [1]{#1}%
\providecommand \enquote  [1]{``#1''}%
\providecommand \bibnamefont  [1]{#1}%
\providecommand \bibfnamefont [1]{#1}%
\providecommand \citenamefont [1]{#1}%
\providecommand \href@noop [0]{\@secondoftwo}%
\providecommand \href [0]{\begingroup \@sanitize@url \@href}%
\providecommand \@href[1]{\@@startlink{#1}\@@href}%
\providecommand \@@href[1]{\endgroup#1\@@endlink}%
\providecommand \@sanitize@url [0]{\catcode `\\12\catcode `\$12\catcode
  `\&12\catcode `\#12\catcode `\^12\catcode `\_12\catcode `\%12\relax}%
\providecommand \@@startlink[1]{}%
\providecommand \@@endlink[0]{}%
\providecommand \url  [0]{\begingroup\@sanitize@url \@url }%
\providecommand \@url [1]{\endgroup\@href {#1}{\urlprefix }}%
\providecommand \urlprefix  [0]{URL }%
\providecommand \Eprint [0]{\href }%
\providecommand \doibase [0]{https://doi.org/}%
\providecommand \selectlanguage [0]{\@gobble}%
\providecommand \bibinfo  [0]{\@secondoftwo}%
\providecommand \bibfield  [0]{\@secondoftwo}%
\providecommand \translation [1]{[#1]}%
\providecommand \BibitemOpen [0]{}%
\providecommand \bibitemStop [0]{}%
\providecommand \bibitemNoStop [0]{.\EOS\space}%
\providecommand \EOS [0]{\spacefactor3000\relax}%
\providecommand \BibitemShut  [1]{\csname bibitem#1\endcsname}%
\let\auto@bib@innerbib\@empty
%</preamble>
\bibitem [{\citenamefont {Giuliani}\ \emph {et~al.}(2019)\citenamefont
  {Giuliani}, \citenamefont {Matheson}, \citenamefont {Nazarewicz},
  \citenamefont {Olsen}, \citenamefont {Reinhard}, \citenamefont {Sadhukhan},
  \citenamefont {Schuetrumpf}, \citenamefont {Schunck},\ and\ \citenamefont
  {Schwerdtfeger}}]{Giuliani2019}%
  \BibitemOpen
  \bibfield  {author} {\bibinfo {author} {\bibfnamefont {S.~A.}\ \bibnamefont
  {Giuliani}}, \bibinfo {author} {\bibfnamefont {Z.}~\bibnamefont {Matheson}},
  \bibinfo {author} {\bibfnamefont {W.}~\bibnamefont {Nazarewicz}}, \bibinfo
  {author} {\bibfnamefont {E.}~\bibnamefont {Olsen}}, \bibinfo {author}
  {\bibfnamefont {P.~G.}\ \bibnamefont {Reinhard}}, \bibinfo {author}
  {\bibfnamefont {J.}~\bibnamefont {Sadhukhan}}, \bibinfo {author}
  {\bibfnamefont {B.}~\bibnamefont {Schuetrumpf}}, \bibinfo {author}
  {\bibfnamefont {N.}~\bibnamefont {Schunck}},\ and\ \bibinfo {author}
  {\bibfnamefont {P.}~\bibnamefont {Schwerdtfeger}},\ }\href
  {https://doi.org/10.1103/RevModPhys.91.011001} {\bibfield  {journal}
  {\bibinfo  {journal} {Rev. Mod. Phys.}\ }\textbf {\bibinfo {volume} {91}},\
  \bibinfo {pages} {11001} (\bibinfo {year} {2019})}\BibitemShut {NoStop}%
\bibitem [{\citenamefont {Hofmann}(2015)}]{Hofmann2015}%
  \BibitemOpen
  \bibfield  {author} {\bibinfo {author} {\bibfnamefont {S.}~\bibnamefont
  {Hofmann}},\ }\href {https://doi.org/10.1088/0954-3899/42/11/114001}
  {\bibfield  {journal} {\bibinfo  {journal} {J. Phys. G Nucl. Part. Phys.}\
  }\textbf {\bibinfo {volume} {42}},\ \bibinfo {pages} {114001} (\bibinfo
  {year} {2015})}\BibitemShut {NoStop}%
\bibitem [{\citenamefont {Oganessian}\ and\ \citenamefont
  {Utyonkov}(2015)}]{Oganessian2015}%
  \BibitemOpen
  \bibfield  {author} {\bibinfo {author} {\bibfnamefont {Y.}~\bibnamefont
  {Oganessian}}\ and\ \bibinfo {author} {\bibfnamefont {V.}~\bibnamefont
  {Utyonkov}},\ }\href {https://doi.org/10.1016/j.nuclphysa.2015.07.003}
  {\bibfield  {journal} {\bibinfo  {journal} {Nucl. Phys. A}\ }\textbf
  {\bibinfo {volume} {944}},\ \bibinfo {pages} {62} (\bibinfo {year}
  {2015})}\BibitemShut {NoStop}%
\bibitem [{\citenamefont {Oganessian}\ \emph {et~al.}(2017)\citenamefont
  {Oganessian}, \citenamefont {Sobiczewski},\ and\ \citenamefont
  {Ter-Akopian}}]{Oganessian2017}%
  \BibitemOpen
  \bibfield  {author} {\bibinfo {author} {\bibfnamefont {Y.~T.}\ \bibnamefont
  {Oganessian}}, \bibinfo {author} {\bibfnamefont {A.}~\bibnamefont
  {Sobiczewski}},\ and\ \bibinfo {author} {\bibfnamefont {G.~M.}\ \bibnamefont
  {Ter-Akopian}},\ }\href {https://doi.org/10.1088/1402-4896/aa53c1} {\bibfield
   {journal} {\bibinfo  {journal} {Phys. Scr.}\ }\textbf {\bibinfo {volume}
  {92}},\ \bibinfo {pages} {023003} (\bibinfo {year} {2017})}\BibitemShut
  {NoStop}%
\bibitem [{\citenamefont {Haba}(2019)}]{haba2019}%
  \BibitemOpen
  \bibfield  {author} {\bibinfo {author} {\bibfnamefont {H.}~\bibnamefont
  {Haba}},\ }\href {https://doi.org/10.1038/s41557-018-0191-8} {\bibfield
  {journal} {\bibinfo  {journal} {Nat. Chem.}\ }\textbf {\bibinfo {volume}
  {11}},\ \bibinfo {pages} {10} (\bibinfo {year} {2019})}\BibitemShut {NoStop}%
\bibitem [{\citenamefont {Ackermann}\ and\ \citenamefont
  {Theisen}(2017)}]{Ackermann2017}%
  \BibitemOpen
  \bibfield  {author} {\bibinfo {author} {\bibfnamefont {D.}~\bibnamefont
  {Ackermann}}\ and\ \bibinfo {author} {\bibfnamefont {C.}~\bibnamefont
  {Theisen}},\ }\href {https://doi.org/10.1088/1402-4896/aa7921} {\bibfield
  {journal} {\bibinfo  {journal} {Phys. Scr.}\ }\textbf {\bibinfo {volume}
  {92}},\ \bibinfo {pages} {083002} (\bibinfo {year} {2017})}\BibitemShut
  {NoStop}%
\bibitem [{\citenamefont {Sridhar}\ \emph {et~al.}(2018)\citenamefont
  {Sridhar}, \citenamefont {Manjunatha},\ and\ \citenamefont
  {Ramalingam}}]{Sridhar2015}%
  \BibitemOpen
  \bibfield  {author} {\bibinfo {author} {\bibfnamefont {K.~N.}\ \bibnamefont
  {Sridhar}}, \bibinfo {author} {\bibfnamefont {H.~C.}\ \bibnamefont
  {Manjunatha}},\ and\ \bibinfo {author} {\bibfnamefont {H.~B.}\ \bibnamefont
  {Ramalingam}},\ }\href {https://doi.org/10.1103/PhysRevC.98.064605}
  {\bibfield  {journal} {\bibinfo  {journal} {Phys. Rev. C}\ }\textbf {\bibinfo
  {volume} {98}},\ \bibinfo {pages} {064605} (\bibinfo {year}
  {2018})}\BibitemShut {NoStop}%
\bibitem [{\citenamefont {Voinov}\ \emph {et~al.}(2020)\citenamefont {Voinov},
  \citenamefont {Utyonkov}, \citenamefont {Oganessian}, \citenamefont
  {Abdullin}, \citenamefont {Polyakov}, \citenamefont {Tsyganov}, \citenamefont
  {Shirokovsky}, \citenamefont {Sagaidak}, \citenamefont {Subbotin},
  \citenamefont {Dmitriev}, \citenamefont {Itkis}, \citenamefont {Shumeiko},
  \citenamefont {Kovrizhnykh}, \citenamefont {Sabelnikov},\ and\ \citenamefont
  {Vostokin}}]{Voinov2020}%
  \BibitemOpen
  \bibfield  {author} {\bibinfo {author} {\bibfnamefont {A.~A.}\ \bibnamefont
  {Voinov}}, \bibinfo {author} {\bibfnamefont {V.~K.}\ \bibnamefont
  {Utyonkov}}, \bibinfo {author} {\bibfnamefont {Y.~T.}\ \bibnamefont
  {Oganessian}}, \bibinfo {author} {\bibfnamefont {F.~S.}\ \bibnamefont
  {Abdullin}}, \bibinfo {author} {\bibfnamefont {A.~N.}\ \bibnamefont
  {Polyakov}}, \bibinfo {author} {\bibfnamefont {Y.~S.}\ \bibnamefont
  {Tsyganov}}, \bibinfo {author} {\bibfnamefont {I.~V.}\ \bibnamefont
  {Shirokovsky}}, \bibinfo {author} {\bibfnamefont {R.~N.}\ \bibnamefont
  {Sagaidak}}, \bibinfo {author} {\bibfnamefont {V.~G.}\ \bibnamefont
  {Subbotin}}, \bibinfo {author} {\bibfnamefont {S.~N.}\ \bibnamefont
  {Dmitriev}}, \bibinfo {author} {\bibfnamefont {M.~G.}\ \bibnamefont {Itkis}},
  \bibinfo {author} {\bibfnamefont {M.~V.}\ \bibnamefont {Shumeiko}}, \bibinfo
  {author} {\bibfnamefont {N.~D.}\ \bibnamefont {Kovrizhnykh}}, \bibinfo
  {author} {\bibfnamefont {A.~V.}\ \bibnamefont {Sabelnikov}},\ and\ \bibinfo
  {author} {\bibfnamefont {G.~K.}\ \bibnamefont {Vostokin}},\ }\href
  {https://doi.org/10.3103/S1062873820040358} {\bibfield  {journal} {\bibinfo
  {journal} {Bull. Russ. Acad. Sci. Phys.}\ }\textbf {\bibinfo {volume} {84}},\
  \bibinfo {pages} {351} (\bibinfo {year} {2020})}\BibitemShut {NoStop}%
\bibitem [{\citenamefont {Kayumov}\ \emph {et~al.}(2022)\citenamefont
  {Kayumov}, \citenamefont {Ganiev}, \citenamefont {Nasirov},\ and\
  \citenamefont {Yuldasheva}}]{kayumov2022}%
  \BibitemOpen
  \bibfield  {author} {\bibinfo {author} {\bibfnamefont {B.~M.}\ \bibnamefont
  {Kayumov}}, \bibinfo {author} {\bibfnamefont {O.~K.}\ \bibnamefont {Ganiev}},
  \bibinfo {author} {\bibfnamefont {A.~K.}\ \bibnamefont {Nasirov}},\ and\
  \bibinfo {author} {\bibfnamefont {G.~A.}\ \bibnamefont {Yuldasheva}},\ }\href
  {https://doi.org/10.1103/PhysRevC.105.014618} {\bibfield  {journal} {\bibinfo
   {journal} {Phys. Rev. C}\ }\textbf {\bibinfo {volume} {105}},\ \bibinfo
  {pages} {014618} (\bibinfo {year} {2022})}\BibitemShut {NoStop}%
\bibitem [{\citenamefont {Tian}\ \emph {et~al.}(2008)\citenamefont {Tian},
  \citenamefont {Wu}, \citenamefont {Zhao}, \citenamefont {Zhang},\ and\
  \citenamefont {Li}}]{Tian2008}%
  \BibitemOpen
  \bibfield  {author} {\bibinfo {author} {\bibfnamefont {J.}~\bibnamefont
  {Tian}}, \bibinfo {author} {\bibfnamefont {X.}~\bibnamefont {Wu}}, \bibinfo
  {author} {\bibfnamefont {K.}~\bibnamefont {Zhao}}, \bibinfo {author}
  {\bibfnamefont {Y.}~\bibnamefont {Zhang}},\ and\ \bibinfo {author}
  {\bibfnamefont {Z.}~\bibnamefont {Li}},\ }\href
  {https://doi.org/10.1103/PhysRevC.77.064603} {\bibfield  {journal} {\bibinfo
  {journal} {Phys. Rev. C}\ }\textbf {\bibinfo {volume} {77}},\ \bibinfo
  {pages} {064603} (\bibinfo {year} {2008})}\BibitemShut {NoStop}%
\bibitem [{\citenamefont {Watanabe}\ \emph {et~al.}(2015)\citenamefont
  {Watanabe}, \citenamefont {Kim}, \citenamefont {Jeong}, \citenamefont
  {Hirayama}, \citenamefont {Imai}, \citenamefont {Ishiyama}, \citenamefont
  {Jung}, \citenamefont {Miyatake}, \citenamefont {Choi}, \citenamefont {Song},
  \citenamefont {Clement}, \citenamefont {de~France}, \citenamefont {Navin},
  \citenamefont {Rejmund}, \citenamefont {Schmitt}, \citenamefont {Pollarolo},
  \citenamefont {Corradi}, \citenamefont {Fioretto}, \citenamefont {Montanari},
  \citenamefont {Niikura}, \citenamefont {Suzuki}, \citenamefont {Nishibata},\
  and\ \citenamefont {Takatsu}}]{Watanabe2015}%
  \BibitemOpen
  \bibfield  {author} {\bibinfo {author} {\bibfnamefont {Y.~X.}\ \bibnamefont
  {Watanabe}}, \bibinfo {author} {\bibfnamefont {Y.~H.}\ \bibnamefont {Kim}},
  \bibinfo {author} {\bibfnamefont {S.~C.}\ \bibnamefont {Jeong}}, \bibinfo
  {author} {\bibfnamefont {Y.}~\bibnamefont {Hirayama}}, \bibinfo {author}
  {\bibfnamefont {N.}~\bibnamefont {Imai}}, \bibinfo {author} {\bibfnamefont
  {H.}~\bibnamefont {Ishiyama}}, \bibinfo {author} {\bibfnamefont {H.~S.}\
  \bibnamefont {Jung}}, \bibinfo {author} {\bibfnamefont {H.}~\bibnamefont
  {Miyatake}}, \bibinfo {author} {\bibfnamefont {S.}~\bibnamefont {Choi}},
  \bibinfo {author} {\bibfnamefont {J.~S.}\ \bibnamefont {Song}}, \bibinfo
  {author} {\bibfnamefont {E.}~\bibnamefont {Clement}}, \bibinfo {author}
  {\bibfnamefont {G.}~\bibnamefont {de~France}}, \bibinfo {author}
  {\bibfnamefont {A.}~\bibnamefont {Navin}}, \bibinfo {author} {\bibfnamefont
  {M.}~\bibnamefont {Rejmund}}, \bibinfo {author} {\bibfnamefont
  {C.}~\bibnamefont {Schmitt}}, \bibinfo {author} {\bibfnamefont
  {G.}~\bibnamefont {Pollarolo}}, \bibinfo {author} {\bibfnamefont
  {L.}~\bibnamefont {Corradi}}, \bibinfo {author} {\bibfnamefont
  {E.}~\bibnamefont {Fioretto}}, \bibinfo {author} {\bibfnamefont
  {D.}~\bibnamefont {Montanari}}, \bibinfo {author} {\bibfnamefont
  {M.}~\bibnamefont {Niikura}}, \bibinfo {author} {\bibfnamefont
  {D.}~\bibnamefont {Suzuki}}, \bibinfo {author} {\bibfnamefont
  {H.}~\bibnamefont {Nishibata}},\ and\ \bibinfo {author} {\bibfnamefont
  {J.}~\bibnamefont {Takatsu}},\ }\href
  {https://doi.org/10.1103/PhysRevLett.115.172503} {\bibfield  {journal}
  {\bibinfo  {journal} {Phys. Rev. Lett.}\ }\textbf {\bibinfo {volume} {115}},\
  \bibinfo {pages} {172503} (\bibinfo {year} {2015})}\BibitemShut {NoStop}%
\bibitem [{\citenamefont {Saiko}\ and\ \citenamefont
  {Karpov}(2022)}]{Saiko2022}%
  \BibitemOpen
  \bibfield  {author} {\bibinfo {author} {\bibfnamefont {V.}~\bibnamefont
  {Saiko}}\ and\ \bibinfo {author} {\bibfnamefont {A.}~\bibnamefont {Karpov}},\
  }\href {https://doi.org/10.1140/epja/s10050-022-00688-9} {\bibfield
  {journal} {\bibinfo  {journal} {Eur. Phys. J. A}\ }\textbf {\bibinfo {volume}
  {58}},\ \bibinfo {pages} {41} (\bibinfo {year} {2022})}\BibitemShut {NoStop}%
\bibitem [{\citenamefont {Zagrebaev}\ and\ \citenamefont
  {Greiner}(2015)}]{zagrebaev2015}%
  \BibitemOpen
  \bibfield  {author} {\bibinfo {author} {\bibfnamefont {V.}~\bibnamefont
  {Zagrebaev}}\ and\ \bibinfo {author} {\bibfnamefont {W.}~\bibnamefont
  {Greiner}},\ }\href {https://doi.org/10.1016/j.nuclphysa.2015.02.010}
  {\bibfield  {journal} {\bibinfo  {journal} {Nucl. Phys. A}\ }\textbf
  {\bibinfo {volume} {944}},\ \bibinfo {pages} {257} (\bibinfo {year}
  {2015})}\BibitemShut {NoStop}%
\bibitem [{\citenamefont {Bass}(1977)}]{Bass1977}%
  \BibitemOpen
  \bibfield  {author} {\bibinfo {author} {\bibfnamefont {R.}~\bibnamefont
  {Bass}},\ }\href {https://doi.org/10.1103/PhysRevLett.39.265} {\bibfield
  {journal} {\bibinfo  {journal} {Phys. Rev. Lett.}\ }\textbf {\bibinfo
  {volume} {39}},\ \bibinfo {pages} {265} (\bibinfo {year} {1977})}\BibitemShut
  {NoStop}%
\bibitem [{\citenamefont {Ghodsi}\ and\ \citenamefont
  {Lari}(2013)}]{GHODSI2013}%
  \BibitemOpen
  \bibfield  {author} {\bibinfo {author} {\bibfnamefont {O.~N.}\ \bibnamefont
  {Ghodsi}}\ and\ \bibinfo {author} {\bibfnamefont {F.}~\bibnamefont {Lari}},\
  }\href {https://doi.org/10.1142/S0217732313501162} {\bibfield  {journal}
  {\bibinfo  {journal} {Mod. Phys. Lett. A}\ }\textbf {\bibinfo {volume}
  {28}},\ \bibinfo {pages} {1350116} (\bibinfo {year} {2013})}\BibitemShut
  {NoStop}%
\bibitem [{\citenamefont {Manjunatha}\ and\ \citenamefont
  {Sridhar}(2018)}]{MANJUNATHA2018}%
  \BibitemOpen
  \bibfield  {author} {\bibinfo {author} {\bibfnamefont {H.}~\bibnamefont
  {Manjunatha}}\ and\ \bibinfo {author} {\bibfnamefont {K.}~\bibnamefont
  {Sridhar}},\ }\href {https://doi.org/10.1016/j.nuclphysa.2018.01.016}
  {\bibfield  {journal} {\bibinfo  {journal} {Nucl. Phys. A}\ }\textbf
  {\bibinfo {volume} {971}},\ \bibinfo {pages} {83} (\bibinfo {year}
  {2018})}\BibitemShut {NoStop}%
\bibitem [{\citenamefont {Zagrebaev}\ \emph
  {et~al.}(2007{\natexlab{a}})\citenamefont {Zagrebaev}, \citenamefont
  {Karpov}, \citenamefont {Aritomo}, \citenamefont {Naumenko},\ and\
  \citenamefont {Greiner}}]{Zagrebaev2007}%
  \BibitemOpen
  \bibfield  {author} {\bibinfo {author} {\bibfnamefont {V.}~\bibnamefont
  {Zagrebaev}}, \bibinfo {author} {\bibfnamefont {A.}~\bibnamefont {Karpov}},
  \bibinfo {author} {\bibfnamefont {Y.}~\bibnamefont {Aritomo}}, \bibinfo
  {author} {\bibfnamefont {M.}~\bibnamefont {Naumenko}},\ and\ \bibinfo
  {author} {\bibfnamefont {W.}~\bibnamefont {Greiner}},\ }\href
  {https://doi.org/10.1134/S106377960704003X} {\bibfield  {journal} {\bibinfo
  {journal} {Phys. Part. Nucl.}\ }\textbf {\bibinfo {volume} {38}},\ \bibinfo
  {pages} {469} (\bibinfo {year} {2007}{\natexlab{a}})}\BibitemShut {NoStop}%
\bibitem [{\citenamefont {Qu}\ \emph {et~al.}(2014)\citenamefont {Qu},
  \citenamefont {Zhang}, \citenamefont {Zhang},\ and\ \citenamefont
  {Wolski}}]{qu2014}%
  \BibitemOpen
  \bibfield  {author} {\bibinfo {author} {\bibfnamefont {W.~W.}\ \bibnamefont
  {Qu}}, \bibinfo {author} {\bibfnamefont {G.~L.}\ \bibnamefont {Zhang}},
  \bibinfo {author} {\bibfnamefont {H.~Q.}\ \bibnamefont {Zhang}},\ and\
  \bibinfo {author} {\bibfnamefont {R.}~\bibnamefont {Wolski}},\ }\href
  {https://doi.org/10.1103/PhysRevC.90.064603} {\bibfield  {journal} {\bibinfo
  {journal} {Phys. Rev. C}\ }\textbf {\bibinfo {volume} {90}},\ \bibinfo
  {pages} {064603} (\bibinfo {year} {2014})}\BibitemShut {NoStop}%
\bibitem [{\citenamefont {Back}\ \emph {et~al.}(2014)\citenamefont {Back},
  \citenamefont {Esbensen}, \citenamefont {Jiang},\ and\ \citenamefont
  {Rehm}}]{Back2014}%
  \BibitemOpen
  \bibfield  {author} {\bibinfo {author} {\bibfnamefont {B.~B.}\ \bibnamefont
  {Back}}, \bibinfo {author} {\bibfnamefont {H.}~\bibnamefont {Esbensen}},
  \bibinfo {author} {\bibfnamefont {C.~L.}\ \bibnamefont {Jiang}},\ and\
  \bibinfo {author} {\bibfnamefont {K.~E.}\ \bibnamefont {Rehm}},\ }\href
  {https://doi.org/10.1103/RevModPhys.86.317} {\bibfield  {journal} {\bibinfo
  {journal} {Rev. Mod. Phys.}\ }\textbf {\bibinfo {volume} {86}},\ \bibinfo
  {pages} {317} (\bibinfo {year} {2014})}\BibitemShut {NoStop}%
\bibitem [{\citenamefont {Zagrebaev}\ \emph
  {et~al.}(2007{\natexlab{b}})\citenamefont {Zagrebaev}, \citenamefont
  {Samarin},\ and\ \citenamefont {Greiner}}]{Zagrebaev2007_astro}%
  \BibitemOpen
  \bibfield  {author} {\bibinfo {author} {\bibfnamefont {V.~I.}\ \bibnamefont
  {Zagrebaev}}, \bibinfo {author} {\bibfnamefont {V.~V.}\ \bibnamefont
  {Samarin}},\ and\ \bibinfo {author} {\bibfnamefont {W.}~\bibnamefont
  {Greiner}},\ }\href {https://doi.org/10.1103/PhysRevC.75.035809} {\bibfield
  {journal} {\bibinfo  {journal} {Phys. Rev. C}\ }\textbf {\bibinfo {volume}
  {75}},\ \bibinfo {pages} {035809} (\bibinfo {year}
  {2007}{\natexlab{b}})}\BibitemShut {NoStop}%
\bibitem [{\citenamefont {Adamian}\ \emph {et~al.}(2016)\citenamefont
  {Adamian}, \citenamefont {Antonenko}, \citenamefont {Lenske}, \citenamefont
  {Tolokonnikov},\ and\ \citenamefont {Saperstein}}]{Adamian2016}%
  \BibitemOpen
  \bibfield  {author} {\bibinfo {author} {\bibfnamefont {G.~G.}\ \bibnamefont
  {Adamian}}, \bibinfo {author} {\bibfnamefont {N.~V.}\ \bibnamefont
  {Antonenko}}, \bibinfo {author} {\bibfnamefont {H.}~\bibnamefont {Lenske}},
  \bibinfo {author} {\bibfnamefont {S.~V.}\ \bibnamefont {Tolokonnikov}},\ and\
  \bibinfo {author} {\bibfnamefont {E.~E.}\ \bibnamefont {Saperstein}},\ }\href
  {https://doi.org/10.1103/PhysRevC.94.054309} {\bibfield  {journal} {\bibinfo
  {journal} {Phys. Rev. C}\ }\textbf {\bibinfo {volume} {94}},\ \bibinfo
  {pages} {054309} (\bibinfo {year} {2016})}\BibitemShut {NoStop}%
\bibitem [{\citenamefont {Sukhareva}\ \emph {et~al.}(2021)\citenamefont
  {Sukhareva}, \citenamefont {Chushnyakova}, \citenamefont {Gontchar},\ and\
  \citenamefont {Klimochkina}}]{Sukhareva2021}%
  \BibitemOpen
  \bibfield  {author} {\bibinfo {author} {\bibfnamefont {O.~M.}\ \bibnamefont
  {Sukhareva}}, \bibinfo {author} {\bibfnamefont {M.~V.}\ \bibnamefont
  {Chushnyakova}}, \bibinfo {author} {\bibfnamefont {I.~I.}\ \bibnamefont
  {Gontchar}},\ and\ \bibinfo {author} {\bibfnamefont {A.~A.}\ \bibnamefont
  {Klimochkina}},\ }\href {https://doi.org/10.3103/S106287382105021X}
  {\bibfield  {journal} {\bibinfo  {journal} {Bull. Russ. Acad. Sci. Phys.}\
  }\textbf {\bibinfo {volume} {85}},\ \bibinfo {pages} {508} (\bibinfo {year}
  {2021})}\BibitemShut {NoStop}%
\bibitem [{\citenamefont {Chamon}\ \emph {et~al.}(2002)\citenamefont {Chamon},
  \citenamefont {Carlson}, \citenamefont {Gasques}, \citenamefont {Pereira},
  \citenamefont {De Conti}, \citenamefont {Alvarez}, \citenamefont {Hussein},
  \citenamefont {Ribeiro}, \citenamefont {Rossi},\ and\ \citenamefont
  {Silva}}]{Chamon2002}%
  \BibitemOpen
  \bibfield  {author} {\bibinfo {author} {\bibfnamefont {L.~C.}\ \bibnamefont
  {Chamon}}, \bibinfo {author} {\bibfnamefont {B.~V.}\ \bibnamefont {Carlson}},
  \bibinfo {author} {\bibfnamefont {L.~R.}\ \bibnamefont {Gasques}}, \bibinfo
  {author} {\bibfnamefont {D.}~\bibnamefont {Pereira}}, \bibinfo {author}
  {\bibfnamefont {C.}~\bibnamefont {De Conti}}, \bibinfo {author} {\bibfnamefont
  {M.~A.~G.}\ \bibnamefont {Alvarez}}, \bibinfo {author} {\bibfnamefont
  {M.~S.}\ \bibnamefont {Hussein}}, \bibinfo {author} {\bibfnamefont
  {M.~A.~C.}\ \bibnamefont {Ribeiro}}, \bibinfo {author} {\bibfnamefont
  {E.~S.}\ \bibnamefont {Rossi}},\ and\ \bibinfo {author} {\bibfnamefont
  {C.~P.}\ \bibnamefont {Silva}},\ }\href
  {https://doi.org/10.1103/PhysRevC.66.014610} {\bibfield  {journal} {\bibinfo
  {journal} {Phys. Rev. C}\ }\textbf {\bibinfo {volume} {66}},\ \bibinfo
  {pages} {014610} (\bibinfo {year} {2002})}\BibitemShut {NoStop}%
\bibitem [{\citenamefont {{De Vries}}\ \emph {et~al.}(1987)\citenamefont {{De
  Vries}}, \citenamefont {{De Jager}},\ and\ \citenamefont {{De
  Vries}}}]{DeVries1987}%
  \BibitemOpen
  \bibfield  {author} {\bibinfo {author} {\bibfnamefont {H.}~\bibnamefont {{De
  Vries}}}, \bibinfo {author} {\bibfnamefont {C.~W.}\ \bibnamefont {{De
  Jager}}},\ and\ \bibinfo {author} {\bibfnamefont {C.}~\bibnamefont {{De
  Vries}}},\ }\href {https://doi.org/10.1016/0092-640X(87)90013-1} {\bibfield
  {journal} {\bibinfo  {journal} {At. Data Nucl. Data Tables}\ }\textbf
  {\bibinfo {volume} {36}},\ \bibinfo {pages} {495} (\bibinfo {year}
  {1987})}\BibitemShut {NoStop}%
\bibitem [{\citenamefont {Hasse}\ and\ \citenamefont
  {Myers}(1988)}]{Hasse1988}%
  \BibitemOpen
  \bibfield  {author} {\bibinfo {author} {\bibfnamefont {R.~W.}\ \bibnamefont
  {Hasse}}\ and\ \bibinfo {author} {\bibfnamefont {W.~D.}\ \bibnamefont
  {Myers}},\ }\href {https://doi.org/10.1007/978-3-642-83017-4} {\emph
  {\bibinfo {title} {Geometrical Relationships of Macroscopic Nuclear
  Physics}}}\ (\bibinfo  {publisher} {Springer Berlin Heidelberg},\ \bibinfo
  {year} {1988})\ pp.\ \bibinfo {pages} {1--114}\BibitemShut {NoStop}%
\bibitem [{\citenamefont {Abdulghany}(2018)}]{Abdulghany2018}%
  \BibitemOpen
  \bibfield  {author} {\bibinfo {author} {\bibfnamefont {A.~R.}\ \bibnamefont
  {Abdulghany}},\ }\href {https://doi.org/10.1088/1674-1137/42/7/074101}
  {\bibfield  {journal} {\bibinfo  {journal} {Chinese Phys. C}\ }\textbf
  {\bibinfo {volume} {42}},\ \bibinfo {pages} {074101} (\bibinfo {year}
  {2018})}\BibitemShut {NoStop}%
\bibitem [{\citenamefont {Friar}\ and\ \citenamefont
  {Negele}(1975)}]{Friar1975}%
  \BibitemOpen
  \bibfield  {author} {\bibinfo {author} {\bibfnamefont {J.~L.}\ \bibnamefont
  {Friar}}\ and\ \bibinfo {author} {\bibfnamefont {J.~W.}\ \bibnamefont
  {Negele}},\ }in\ \href {https://doi.org/10.1007/978-1-4757-4398-2\_3} {\emph
  {\bibinfo {booktitle} {Advances in Nuclear Physics}}}\ (\bibinfo  {publisher}
  {Springer US},\ \bibinfo {address} {Boston, MA},\ \bibinfo {year} {1975})\
  pp.\ \bibinfo {pages} {219--376}\BibitemShut {NoStop}%
\bibitem [{\citenamefont {Zyla}\ \emph {et~al.}(2020)\citenamefont {Zyla} \emph
  {et~al.}}]{Zyla2020_PDG}%
  \BibitemOpen
  \bibfield  {author} {\bibinfo {author} {\bibfnamefont {P.}~\bibnamefont
  {Zyla}} \emph {et~al.} (\bibinfo {collaboration} {Particle Data Group}),\
  }\href {https://doi.org/10.1093/ptep/ptaa104} {\bibfield  {journal} {\bibinfo
   {journal} {Prog. Theor. Exp. Phys.}\ }\textbf {\bibinfo {volume} {2020}},\
  \bibinfo {pages} {083C01} (\bibinfo {year} {2020})}\BibitemShut {NoStop}%
\bibitem [{\citenamefont {Mohr}\ \emph {et~al.}(2016)\citenamefont {Mohr},
  \citenamefont {Newell},\ and\ \citenamefont {Taylor}}]{CODATA2014}%
  \BibitemOpen
  \bibfield  {author} {\bibinfo {author} {\bibfnamefont {P.~J.}\ \bibnamefont
  {Mohr}}, \bibinfo {author} {\bibfnamefont {D.~B.}\ \bibnamefont {Newell}},\
  and\ \bibinfo {author} {\bibfnamefont {B.~N.}\ \bibnamefont {Taylor}},\
  }\href {https://doi.org/10.1063/1.4954402} {\bibfield  {journal} {\bibinfo
  {journal} {J. Phys. Chem. Ref. Data}\ }\textbf {\bibinfo {volume} {45}},\
  \bibinfo {pages} {043102} (\bibinfo {year} {2016})}\BibitemShut {NoStop}%
\bibitem [{\citenamefont {Tiesinga}\ \emph {et~al.}(2021)\citenamefont
  {Tiesinga}, \citenamefont {Mohr}, \citenamefont {Newell},\ and\ \citenamefont
  {Taylor}}]{CODATA2018}%
  \BibitemOpen
  \bibfield  {author} {\bibinfo {author} {\bibfnamefont {E.}~\bibnamefont
  {Tiesinga}}, \bibinfo {author} {\bibfnamefont {P.~J.}\ \bibnamefont {Mohr}},
  \bibinfo {author} {\bibfnamefont {D.~B.}\ \bibnamefont {Newell}},\ and\
  \bibinfo {author} {\bibfnamefont {B.~N.}\ \bibnamefont {Taylor}},\ }\href
  {https://doi.org/10.1103/RevModPhys.93.025010} {\bibfield  {journal}
  {\bibinfo  {journal} {Rev. Mod. Phys.}\ }\textbf {\bibinfo {volume} {93}},\
  \bibinfo {pages} {025010} (\bibinfo {year} {2021})}\BibitemShut {NoStop}%
\bibitem [{\citenamefont {Karr}\ \emph {et~al.}(2020)\citenamefont {Karr},
  \citenamefont {Marchand},\ and\ \citenamefont {Voutier}}]{Karr2020}%
  \BibitemOpen
  \bibfield  {author} {\bibinfo {author} {\bibfnamefont {J.-P.}\ \bibnamefont
  {Karr}}, \bibinfo {author} {\bibfnamefont {D.}~\bibnamefont {Marchand}},\
  and\ \bibinfo {author} {\bibfnamefont {E.}~\bibnamefont {Voutier}},\ }\href
  {https://doi.org/10.1038/s42254-020-0229-x} {\bibfield  {journal} {\bibinfo
  {journal} {Nat. Rev. Phys.}\ }\textbf {\bibinfo {volume} {2}},\ \bibinfo
  {pages} {601} (\bibinfo {year} {2020})}\BibitemShut {NoStop}%
\bibitem [{\citenamefont {Gao}\ and\ \citenamefont
  {Vanderhaeghen}(2022)}]{Gao2022}%
  \BibitemOpen
  \bibfield  {author} {\bibinfo {author} {\bibfnamefont {H.}~\bibnamefont
  {Gao}}\ and\ \bibinfo {author} {\bibfnamefont {M.}~\bibnamefont
  {Vanderhaeghen}},\ }\href {https://doi.org/10.1103/RevModPhys.94.015002}
  {\bibfield  {journal} {\bibinfo  {journal} {Rev. Mod. Phys.}\ }\textbf
  {\bibinfo {volume} {94}},\ \bibinfo {pages} {015002} (\bibinfo {year}
  {2022})}\BibitemShut {NoStop}%
\bibitem [{\citenamefont {Lima}\ \emph {et~al.}(2004)\citenamefont {Lima},
  \citenamefont {L{\'{e}}pine-Szily}, \citenamefont {Villari}, \citenamefont
  {Mittig}, \citenamefont {Lichtenth{\"{a}}ler}, \citenamefont {Chartier},
  \citenamefont {Orr}, \citenamefont {Ang{\'{e}}lique}, \citenamefont {Audi},
  \citenamefont {Baldini-Neto}, \citenamefont {Carlson}, \citenamefont
  {Casandjian}, \citenamefont {Cunsolo}, \citenamefont {Donzaud}, \citenamefont
  {Foti}, \citenamefont {Gillibert}, \citenamefont {Hirata}, \citenamefont
  {Lewitowicz}, \citenamefont {Lukyanov}, \citenamefont {MacCormick},
  \citenamefont {Morrissey}, \citenamefont {Ostrowski}, \citenamefont
  {Sherrill}, \citenamefont {Stephan}, \citenamefont {Suomij{\"{a}}rvi},
  \citenamefont {Tassan-Got}, \citenamefont {Vieira},\ and\ \citenamefont
  {Wouters}}]{Lima2004}%
  \BibitemOpen
  \bibfield  {author} {\bibinfo {author} {\bibfnamefont {G.~F.}\ \bibnamefont
  {Lima}}, \bibinfo {author} {\bibfnamefont {A.}~\bibnamefont
  {L{\'{e}}pine-Szily}}, \bibinfo {author} {\bibfnamefont {A.~C.}\ \bibnamefont
  {Villari}}, \bibinfo {author} {\bibfnamefont {W.}~\bibnamefont {Mittig}},
  \bibinfo {author} {\bibfnamefont {R.}~\bibnamefont {Lichtenth{\"{a}}ler}},
  \bibinfo {author} {\bibfnamefont {M.}~\bibnamefont {Chartier}}, \bibinfo
  {author} {\bibfnamefont {N.~A.}\ \bibnamefont {Orr}}, \bibinfo {author}
  {\bibfnamefont {J.~C.}\ \bibnamefont {Ang{\'{e}}lique}}, \bibinfo {author}
  {\bibfnamefont {G.}~\bibnamefont {Audi}}, \bibinfo {author} {\bibfnamefont
  {E.}~\bibnamefont {Baldini-Neto}}, \bibinfo {author} {\bibfnamefont {B.~V.}\
  \bibnamefont {Carlson}}, \bibinfo {author} {\bibfnamefont {J.~M.}\
  \bibnamefont {Casandjian}}, \bibinfo {author} {\bibfnamefont
  {A.}~\bibnamefont {Cunsolo}}, \bibinfo {author} {\bibfnamefont
  {C.}~\bibnamefont {Donzaud}}, \bibinfo {author} {\bibfnamefont
  {A.}~\bibnamefont {Foti}}, \bibinfo {author} {\bibfnamefont {A.}~\bibnamefont
  {Gillibert}}, \bibinfo {author} {\bibfnamefont {D.}~\bibnamefont {Hirata}},
  \bibinfo {author} {\bibfnamefont {M.}~\bibnamefont {Lewitowicz}}, \bibinfo
  {author} {\bibfnamefont {S.}~\bibnamefont {Lukyanov}}, \bibinfo {author}
  {\bibfnamefont {M.}~\bibnamefont {MacCormick}}, \bibinfo {author}
  {\bibfnamefont {D.~J.}\ \bibnamefont {Morrissey}}, \bibinfo {author}
  {\bibfnamefont {A.~N.}\ \bibnamefont {Ostrowski}}, \bibinfo {author}
  {\bibfnamefont {B.~M.}\ \bibnamefont {Sherrill}}, \bibinfo {author}
  {\bibfnamefont {C.}~\bibnamefont {Stephan}}, \bibinfo {author} {\bibfnamefont
  {T.}~\bibnamefont {Suomij{\"{a}}rvi}}, \bibinfo {author} {\bibfnamefont
  {L.}~\bibnamefont {Tassan-Got}}, \bibinfo {author} {\bibfnamefont {D.~J.}\
  \bibnamefont {Vieira}},\ and\ \bibinfo {author} {\bibfnamefont {J.~M.}\
  \bibnamefont {Wouters}},\ }\href
  {https://doi.org/10.1016/j.nuclphysa.2004.01.125} {\bibfield  {journal}
  {\bibinfo  {journal} {Nucl. Phys. A}\ }\textbf {\bibinfo {volume} {735}},\
  \bibinfo {pages} {303} (\bibinfo {year} {2004})}\BibitemShut {NoStop}%
\bibitem [{\citenamefont {Angeli}\ and\ \citenamefont
  {Marinova}(2013)}]{Angeli2013}%
  \BibitemOpen
  \bibfield  {author} {\bibinfo {author} {\bibfnamefont {I.}~\bibnamefont
  {Angeli}}\ and\ \bibinfo {author} {\bibfnamefont {K.~P.}\ \bibnamefont
  {Marinova}},\ }\href {https://doi.org/10.1016/j.adt.2011.12.006} {\bibfield
  {journal} {\bibinfo  {journal} {At. Data Nucl. Data Tables}\ }\textbf
  {\bibinfo {volume} {99}},\ \bibinfo {pages} {69} (\bibinfo {year}
  {2013})}\BibitemShut {NoStop}%
\bibitem [{\citenamefont {Ficenec}\ \emph {et~al.}(1970)\citenamefont
  {Ficenec}, \citenamefont {Trower}, \citenamefont {Heisenberg},\ and\
  \citenamefont {Sick}}]{Ficenec1970}%
  \BibitemOpen
  \bibfield  {author} {\bibinfo {author} {\bibfnamefont {J.}~\bibnamefont
  {Ficenec}}, \bibinfo {author} {\bibfnamefont {W.}~\bibnamefont {Trower}},
  \bibinfo {author} {\bibfnamefont {J.}~\bibnamefont {Heisenberg}},\ and\
  \bibinfo {author} {\bibfnamefont {I.}~\bibnamefont {Sick}},\ }\href
  {https://doi.org/10.1016/0370-2693(70)90383-7} {\bibfield  {journal}
  {\bibinfo  {journal} {Phys. Lett. B}\ }\textbf {\bibinfo {volume} {32}},\
  \bibinfo {pages} {460} (\bibinfo {year} {1970})}\BibitemShut {NoStop}%
\bibitem [{\citenamefont {Wohlfahrt}\ \emph {et~al.}(1980)\citenamefont
  {Wohlfahrt}, \citenamefont {Schwentker}, \citenamefont {Fricke},
  \citenamefont {Andresen},\ and\ \citenamefont {Shera}}]{Wohlfahrt1980}%
  \BibitemOpen
  \bibfield  {author} {\bibinfo {author} {\bibfnamefont {H.~D.}\ \bibnamefont
  {Wohlfahrt}}, \bibinfo {author} {\bibfnamefont {O.}~\bibnamefont
  {Schwentker}}, \bibinfo {author} {\bibfnamefont {G.}~\bibnamefont {Fricke}},
  \bibinfo {author} {\bibfnamefont {H.~G.}\ \bibnamefont {Andresen}},\ and\
  \bibinfo {author} {\bibfnamefont {E.~B.}\ \bibnamefont {Shera}},\ }\href
  {https://doi.org/10.1103/PhysRevC.22.264} {\bibfield  {journal} {\bibinfo
  {journal} {Phys. Rev. C}\ }\textbf {\bibinfo {volume} {22}},\ \bibinfo
  {pages} {264} (\bibinfo {year} {1980})}\BibitemShut {NoStop}%
\bibitem [{\citenamefont {Hammen}\ \emph {et~al.}(2018)\citenamefont {Hammen},
  \citenamefont {N{\"{o}}rtersh{\"{a}}user}, \citenamefont {Balabanski},
  \citenamefont {Bissell}, \citenamefont {Blaum}, \citenamefont
  {Budin{\v{c}}evi{\'{c}}}, \citenamefont {Cheal}, \citenamefont {Flanagan},
  \citenamefont {Fr{\"{o}}mmgen}, \citenamefont {Georgiev}, \citenamefont
  {Geppert}, \citenamefont {Kowalska}, \citenamefont {Kreim}, \citenamefont
  {Krieger}, \citenamefont {Nazarewicz}, \citenamefont {Neugart}, \citenamefont
  {Neyens}, \citenamefont {Papuga}, \citenamefont {Reinhard}, \citenamefont
  {Rajabali}, \citenamefont {Schmidt},\ and\ \citenamefont
  {Yordanov}}]{Hammen2018}%
  \BibitemOpen
  \bibfield  {author} {\bibinfo {author} {\bibfnamefont {M.}~\bibnamefont
  {Hammen}}, \bibinfo {author} {\bibfnamefont {W.}~\bibnamefont
  {N{\"{o}}rtersh{\"{a}}user}}, \bibinfo {author} {\bibfnamefont {D.~L.}\
  \bibnamefont {Balabanski}}, \bibinfo {author} {\bibfnamefont {M.~L.}\
  \bibnamefont {Bissell}}, \bibinfo {author} {\bibfnamefont {K.}~\bibnamefont
  {Blaum}}, \bibinfo {author} {\bibfnamefont {I.}~\bibnamefont
  {Budin{\v{c}}evi{\'{c}}}}, \bibinfo {author} {\bibfnamefont {B.}~\bibnamefont
  {Cheal}}, \bibinfo {author} {\bibfnamefont {K.~T.}\ \bibnamefont {Flanagan}},
  \bibinfo {author} {\bibfnamefont {N.}~\bibnamefont {Fr{\"{o}}mmgen}},
  \bibinfo {author} {\bibfnamefont {G.}~\bibnamefont {Georgiev}}, \bibinfo
  {author} {\bibfnamefont {C.}~\bibnamefont {Geppert}}, \bibinfo {author}
  {\bibfnamefont {M.}~\bibnamefont {Kowalska}}, \bibinfo {author}
  {\bibfnamefont {K.}~\bibnamefont {Kreim}}, \bibinfo {author} {\bibfnamefont
  {A.}~\bibnamefont {Krieger}}, \bibinfo {author} {\bibfnamefont
  {W.}~\bibnamefont {Nazarewicz}}, \bibinfo {author} {\bibfnamefont
  {R.}~\bibnamefont {Neugart}}, \bibinfo {author} {\bibfnamefont
  {G.}~\bibnamefont {Neyens}}, \bibinfo {author} {\bibfnamefont
  {J.}~\bibnamefont {Papuga}}, \bibinfo {author} {\bibfnamefont {P.~G.}\
  \bibnamefont {Reinhard}}, \bibinfo {author} {\bibfnamefont {M.~M.}\
  \bibnamefont {Rajabali}}, \bibinfo {author} {\bibfnamefont {S.}~\bibnamefont
  {Schmidt}},\ and\ \bibinfo {author} {\bibfnamefont {D.~T.}\ \bibnamefont
  {Yordanov}},\ }\href {https://doi.org/10.1103/PhysRevLett.121.102501}
  {\bibfield  {journal} {\bibinfo  {journal} {Phys. Rev. Lett.}\ }\textbf
  {\bibinfo {volume} {121}},\ \bibinfo {pages} {102501} (\bibinfo {year}
  {2018})}\BibitemShut {NoStop}%
\bibitem [{\citenamefont {de~Groote}\ \emph {et~al.}(2020)\citenamefont
  {de~Groote}, \citenamefont {Billowes}, \citenamefont {Binnersley},
  \citenamefont {Bissell}, \citenamefont {Cocolios}, \citenamefont {{Day
  Goodacre}}, \citenamefont {Farooq-Smith}, \citenamefont {Fedorov},
  \citenamefont {Flanagan}, \citenamefont {Franchoo}, \citenamefont {{Garcia
  Ruiz}}, \citenamefont {Gins}, \citenamefont {Holt}, \citenamefont
  {Koszor{\'{u}}s}, \citenamefont {Lynch}, \citenamefont {Miyagi},
  \citenamefont {Nazarewicz}, \citenamefont {Neyens}, \citenamefont {Reinhard},
  \citenamefont {Rothe}, \citenamefont {Stroke}, \citenamefont {Vernon},
  \citenamefont {Wendt}, \citenamefont {Wilkins}, \citenamefont {Xu},\ and\
  \citenamefont {Yang}}]{DeGroote2020}%
  \BibitemOpen
  \bibfield  {author} {\bibinfo {author} {\bibfnamefont {R.~P.}\ \bibnamefont
  {de~Groote}}, \bibinfo {author} {\bibfnamefont {J.}~\bibnamefont {Billowes}},
  \bibinfo {author} {\bibfnamefont {C.~L.}\ \bibnamefont {Binnersley}},
  \bibinfo {author} {\bibfnamefont {M.~L.}\ \bibnamefont {Bissell}}, \bibinfo
  {author} {\bibfnamefont {T.~E.}\ \bibnamefont {Cocolios}}, \bibinfo {author}
  {\bibfnamefont {T.}~\bibnamefont {{Day Goodacre}}}, \bibinfo {author}
  {\bibfnamefont {G.~J.}\ \bibnamefont {Farooq-Smith}}, \bibinfo {author}
  {\bibfnamefont {D.~V.}\ \bibnamefont {Fedorov}}, \bibinfo {author}
  {\bibfnamefont {K.~T.}\ \bibnamefont {Flanagan}}, \bibinfo {author}
  {\bibfnamefont {S.}~\bibnamefont {Franchoo}}, \bibinfo {author}
  {\bibfnamefont {R.~F.}\ \bibnamefont {{Garcia Ruiz}}}, \bibinfo {author}
  {\bibfnamefont {W.}~\bibnamefont {Gins}}, \bibinfo {author} {\bibfnamefont
  {J.~D.}\ \bibnamefont {Holt}}, \bibinfo {author} {\bibfnamefont
  {{\'{A}}.}~\bibnamefont {Koszor{\'{u}}s}}, \bibinfo {author} {\bibfnamefont
  {K.~M.}\ \bibnamefont {Lynch}}, \bibinfo {author} {\bibfnamefont
  {T.}~\bibnamefont {Miyagi}}, \bibinfo {author} {\bibfnamefont
  {W.}~\bibnamefont {Nazarewicz}}, \bibinfo {author} {\bibfnamefont
  {G.}~\bibnamefont {Neyens}}, \bibinfo {author} {\bibfnamefont {P.-G.}\
  \bibnamefont {Reinhard}}, \bibinfo {author} {\bibfnamefont {S.}~\bibnamefont
  {Rothe}}, \bibinfo {author} {\bibfnamefont {H.~H.}\ \bibnamefont {Stroke}},
  \bibinfo {author} {\bibfnamefont {A.~R.}\ \bibnamefont {Vernon}}, \bibinfo
  {author} {\bibfnamefont {K.~D.~A.}\ \bibnamefont {Wendt}}, \bibinfo {author}
  {\bibfnamefont {S.~G.}\ \bibnamefont {Wilkins}}, \bibinfo {author}
  {\bibfnamefont {Z.~Y.}\ \bibnamefont {Xu}},\ and\ \bibinfo {author}
  {\bibfnamefont {X.~F.}\ \bibnamefont {Yang}},\ }\href
  {https://doi.org/10.1038/s41567-020-0868-y} {\bibfield  {journal} {\bibinfo
  {journal} {Nat. Phys.}\ }\textbf {\bibinfo {volume} {16}},\ \bibinfo {pages}
  {620} (\bibinfo {year} {2020})}\BibitemShut {NoStop}%
\bibitem [{\citenamefont {Bayram}\ \emph {et~al.}(2013)\citenamefont {Bayram},
  \citenamefont {Akkoyun}, \citenamefont {Kara},\ and\ \citenamefont
  {Sinan}}]{Bayram2013}%
  \BibitemOpen
  \bibfield  {author} {\bibinfo {author} {\bibfnamefont {T.}~\bibnamefont
  {Bayram}}, \bibinfo {author} {\bibfnamefont {S.}~\bibnamefont {Akkoyun}},
  \bibinfo {author} {\bibfnamefont {S.~O.}\ \bibnamefont {Kara}},\ and\
  \bibinfo {author} {\bibfnamefont {A.}~\bibnamefont {Sinan}},\ }\href
  {https://doi.org/10.5506/APhysPolB.44.1791} {\bibfield  {journal} {\bibinfo
  {journal} {Acta Phys. Pol. B}\ }\textbf {\bibinfo {volume} {44}},\ \bibinfo
  {pages} {1791} (\bibinfo {year} {2013})}\BibitemShut {NoStop}%
\bibitem [{\citenamefont {Angeli}(2004)}]{Angeli2004}%
  \BibitemOpen
  \bibfield  {author} {\bibinfo {author} {\bibfnamefont {I.}~\bibnamefont
  {Angeli}},\ }\href {https://doi.org/10.1016/j.adt.2004.04.002} {\bibfield
  {journal} {\bibinfo  {journal} {At. Data Nucl. Data Tables}\ }\textbf
  {\bibinfo {volume} {87}},\ \bibinfo {pages} {185} (\bibinfo {year}
  {2004})}\BibitemShut {NoStop}%
\bibitem [{\citenamefont {Nerlo-Pomorska}\ and\ \citenamefont
  {Pomorski}(1994)}]{Nerlo-Pomorska1994}%
  \BibitemOpen
  \bibfield  {author} {\bibinfo {author} {\bibfnamefont {B.}~\bibnamefont
  {Nerlo-Pomorska}}\ and\ \bibinfo {author} {\bibfnamefont {K.}~\bibnamefont
  {Pomorski}},\ }\href {https://doi.org/10.1007/BF01291913} {\bibfield
  {journal} {\bibinfo  {journal} {Z. Phys. A}\ }\textbf {\bibinfo {volume}
  {348}},\ \bibinfo {pages} {169} (\bibinfo {year} {1994})} \BibitemShut {NoStop}%
\bibitem [{\citenamefont {Fricke}\ \emph {et~al.}(1995)\citenamefont {Fricke},
  \citenamefont {Bernhardt}, \citenamefont {Heilig}, \citenamefont {Schaller},
  \citenamefont {Schellenberg}, \citenamefont {Shera},\ and\ \citenamefont
  {Dejager}}]{Fricke1995}%
  \BibitemOpen
  \bibfield  {author} {\bibinfo {author} {\bibfnamefont {G.}~\bibnamefont
  {Fricke}}, \bibinfo {author} {\bibfnamefont {C.}~\bibnamefont {Bernhardt}},
  \bibinfo {author} {\bibfnamefont {K.}~\bibnamefont {Heilig}}, \bibinfo
  {author} {\bibfnamefont {L.}~\bibnamefont {Schaller}}, \bibinfo {author}
  {\bibfnamefont {L.}~\bibnamefont {Schellenberg}}, \bibinfo {author}
  {\bibfnamefont {E.}~\bibnamefont {Shera}},\ and\ \bibinfo {author}
  {\bibfnamefont {C.}~\bibnamefont {Dejager}},\ }\href
  {https://doi.org/10.1006/adnd.1995.1007} {\bibfield  {journal} {\bibinfo
  {journal} {At. Data Nucl. Data Tables}\ }\textbf {\bibinfo {volume} {60}},\
  \bibinfo {pages} {177} (\bibinfo {year} {1995})}\BibitemShut {NoStop}%
\bibitem [{\citenamefont {Horowitz}\ \emph {et~al.}(2012)\citenamefont
  {Horowitz}, \citenamefont {Ahmed}, \citenamefont {Jen}, \citenamefont
  {Rakhman}, \citenamefont {Souder}, \citenamefont {Dalton}, \citenamefont
  {Liyanage}, \citenamefont {Paschke}, \citenamefont {Saenboonruang},
  \citenamefont {Silwal}, \citenamefont {Franklin}, \citenamefont {Friend},
  \citenamefont {Quinn}, \citenamefont {Kumar}, \citenamefont {McNulty},
  \citenamefont {Mercado}, \citenamefont {Riordan}, \citenamefont {Wexler},
  \citenamefont {Michaels},\ and\ \citenamefont {Urciuoli}}]{Horowitz2012}%
  \BibitemOpen
  \bibfield  {author} {\bibinfo {author} {\bibfnamefont {C.~J.}\ \bibnamefont
  {Horowitz}}, \bibinfo {author} {\bibfnamefont {Z.}~\bibnamefont {Ahmed}},
  \bibinfo {author} {\bibfnamefont {C.~M.}\ \bibnamefont {Jen}}, \bibinfo
  {author} {\bibfnamefont {A.}~\bibnamefont {Rakhman}}, \bibinfo {author}
  {\bibfnamefont {P.~A.}\ \bibnamefont {Souder}}, \bibinfo {author}
  {\bibfnamefont {M.~M.}\ \bibnamefont {Dalton}}, \bibinfo {author}
  {\bibfnamefont {N.}~\bibnamefont {Liyanage}}, \bibinfo {author}
  {\bibfnamefont {K.~D.}\ \bibnamefont {Paschke}}, \bibinfo {author}
  {\bibfnamefont {K.}~\bibnamefont {Saenboonruang}}, \bibinfo {author}
  {\bibfnamefont {R.}~\bibnamefont {Silwal}}, \bibinfo {author} {\bibfnamefont
  {G.~B.}\ \bibnamefont {Franklin}}, \bibinfo {author} {\bibfnamefont
  {M.}~\bibnamefont {Friend}}, \bibinfo {author} {\bibfnamefont
  {B.}~\bibnamefont {Quinn}}, \bibinfo {author} {\bibfnamefont {K.~S.}\
  \bibnamefont {Kumar}}, \bibinfo {author} {\bibfnamefont {D.}~\bibnamefont
  {McNulty}}, \bibinfo {author} {\bibfnamefont {L.}~\bibnamefont {Mercado}},
  \bibinfo {author} {\bibfnamefont {S.}~\bibnamefont {Riordan}}, \bibinfo
  {author} {\bibfnamefont {J.}~\bibnamefont {Wexler}}, \bibinfo {author}
  {\bibfnamefont {R.~W.}\ \bibnamefont {Michaels}},\ and\ \bibinfo {author}
  {\bibfnamefont {G.~M.}\ \bibnamefont {Urciuoli}},\ }\href
  {https://doi.org/10.1103/PhysRevC.85.032501} {\bibfield  {journal} {\bibinfo
  {journal} {Phys. Rev. C}\ }\textbf {\bibinfo {volume} {85}},\ \bibinfo
  {pages} {032501(R)} (\bibinfo {year} {2012})}\BibitemShut {NoStop}%
\bibitem [{\citenamefont {Krasznahorkay}\ \emph {et~al.}(1999)\citenamefont
  {Krasznahorkay}, \citenamefont {Fujiwara}, \citenamefont {van Aarle},
  \citenamefont {Akimune}, \citenamefont {Daito}, \citenamefont {Fujimura},
  \citenamefont {Fujita}, \citenamefont {Harakeh}, \citenamefont {Inomata},
  \citenamefont {J{\"{a}}necke}, \citenamefont {Nakayama}, \citenamefont
  {Tamii}, \citenamefont {Tanaka}, \citenamefont {Toyokawa}, \citenamefont
  {Uijen},\ and\ \citenamefont {Yosoi}}]{Krasznahorkay1999}%
  \BibitemOpen
  \bibfield  {author} {\bibinfo {author} {\bibfnamefont {A.}~\bibnamefont
  {Krasznahorkay}}, \bibinfo {author} {\bibfnamefont {M.}~\bibnamefont
  {Fujiwara}}, \bibinfo {author} {\bibfnamefont {P.}~\bibnamefont {van Aarle}},
  \bibinfo {author} {\bibfnamefont {H.}~\bibnamefont {Akimune}}, \bibinfo
  {author} {\bibfnamefont {I.}~\bibnamefont {Daito}}, \bibinfo {author}
  {\bibfnamefont {H.}~\bibnamefont {Fujimura}}, \bibinfo {author}
  {\bibfnamefont {Y.}~\bibnamefont {Fujita}}, \bibinfo {author} {\bibfnamefont
  {M.~N.}\ \bibnamefont {Harakeh}}, \bibinfo {author} {\bibfnamefont
  {T.}~\bibnamefont {Inomata}}, \bibinfo {author} {\bibfnamefont
  {J.}~\bibnamefont {J{\"{a}}necke}}, \bibinfo {author} {\bibfnamefont
  {S.}~\bibnamefont {Nakayama}}, \bibinfo {author} {\bibfnamefont
  {A.}~\bibnamefont {Tamii}}, \bibinfo {author} {\bibfnamefont
  {M.}~\bibnamefont {Tanaka}}, \bibinfo {author} {\bibfnamefont
  {H.}~\bibnamefont {Toyokawa}}, \bibinfo {author} {\bibfnamefont
  {W.}~\bibnamefont {Uijen}},\ and\ \bibinfo {author} {\bibfnamefont
  {M.}~\bibnamefont {Yosoi}},\ }\href
  {https://doi.org/10.1103/PhysRevLett.82.3216} {\bibfield  {journal} {\bibinfo
   {journal} {Phys. Rev. Lett.}\ }\textbf {\bibinfo {volume} {82}},\ \bibinfo
  {pages} {3216} (\bibinfo {year} {1999})}\BibitemShut {NoStop}%
\bibitem [{\citenamefont {Klimkiewicz}\ \emph {et~al.}(2007)\citenamefont
  {Klimkiewicz}, \citenamefont {Paar}, \citenamefont {Adrich}, \citenamefont
  {Fallot}, \citenamefont {Boretzky}, \citenamefont {Aumann}, \citenamefont
  {Cortina-Gil}, \citenamefont {Pramanik}, \citenamefont {Elze}, \citenamefont
  {Emling}, \citenamefont {Geissel}, \citenamefont {Hellstr{\"{o}}m},
  \citenamefont {Jones}, \citenamefont {Kratz}, \citenamefont {Kulessa},
  \citenamefont {Nociforo}, \citenamefont {Palit}, \citenamefont {Simon},
  \citenamefont {Sur{\'{o}}wka}, \citenamefont {S{\"{u}}mmerer}, \citenamefont
  {Vretenar},\ and\ \citenamefont {Walu{\'{s}}}}]{Klimkiewicz2007}%
  \BibitemOpen
  \bibfield  {author} {\bibinfo {author} {\bibfnamefont {A.}~\bibnamefont
  {Klimkiewicz}}, \bibinfo {author} {\bibfnamefont {N.}~\bibnamefont {Paar}},
  \bibinfo {author} {\bibfnamefont {P.}~\bibnamefont {Adrich}}, \bibinfo
  {author} {\bibfnamefont {M.}~\bibnamefont {Fallot}}, \bibinfo {author}
  {\bibfnamefont {K.}~\bibnamefont {Boretzky}}, \bibinfo {author}
  {\bibfnamefont {T.}~\bibnamefont {Aumann}}, \bibinfo {author} {\bibfnamefont
  {D.}~\bibnamefont {Cortina-Gil}}, \bibinfo {author} {\bibfnamefont {U.~D.}\
  \bibnamefont {Pramanik}}, \bibinfo {author} {\bibfnamefont {T.~W.}\
  \bibnamefont {Elze}}, \bibinfo {author} {\bibfnamefont {H.}~\bibnamefont
  {Emling}}, \bibinfo {author} {\bibfnamefont {H.}~\bibnamefont {Geissel}},
  \bibinfo {author} {\bibfnamefont {M.}~\bibnamefont {Hellstr{\"{o}}m}},
  \bibinfo {author} {\bibfnamefont {K.~L.}\ \bibnamefont {Jones}}, \bibinfo
  {author} {\bibfnamefont {J.~V.}\ \bibnamefont {Kratz}}, \bibinfo {author}
  {\bibfnamefont {R.}~\bibnamefont {Kulessa}}, \bibinfo {author} {\bibfnamefont
  {C.}~\bibnamefont {Nociforo}}, \bibinfo {author} {\bibfnamefont
  {R.}~\bibnamefont {Palit}}, \bibinfo {author} {\bibfnamefont
  {H.}~\bibnamefont {Simon}}, \bibinfo {author} {\bibfnamefont
  {G.}~\bibnamefont {Sur{\'{o}}wka}}, \bibinfo {author} {\bibfnamefont
  {K.}~\bibnamefont {S{\"{u}}mmerer}}, \bibinfo {author} {\bibfnamefont
  {D.}~\bibnamefont {Vretenar}},\ and\ \bibinfo {author} {\bibfnamefont
  {W.}~\bibnamefont {Walu{\'{s}}}},\ }\href
  {https://doi.org/10.1103/PhysRevC.76.051603} {\bibfield  {journal} {\bibinfo
  {journal} {Phys. Rev. C}\ }\textbf {\bibinfo {volume} {76}},\ \bibinfo
  {pages} {051603(R)} (\bibinfo {year} {2007})}\BibitemShut {NoStop}%
\bibitem [{\citenamefont {Rossi}\ \emph {et~al.}(2013)\citenamefont {Rossi},
  \citenamefont {Adrich}, \citenamefont {Aksouh}, \citenamefont {Alvarez-Pol},
  \citenamefont {Aumann}, \citenamefont {Benlliure}, \citenamefont
  {B{\"{o}}hmer}, \citenamefont {Boretzky}, \citenamefont {Casarejos},
  \citenamefont {Chartier}, \citenamefont {Chatillon}, \citenamefont
  {Cortina-Gil}, \citenamefont {{Datta Pramanik}}, \citenamefont {Emling},
  \citenamefont {Ershova}, \citenamefont {Fernandez-Dominguez}, \citenamefont
  {Geissel}, \citenamefont {Gorska}, \citenamefont {Heil}, \citenamefont
  {Johansson}, \citenamefont {Junghans}, \citenamefont {Kelic-Heil},
  \citenamefont {Kiselev}, \citenamefont {Klimkiewicz}, \citenamefont {Kratz},
  \citenamefont {Kr{\"{u}}cken}, \citenamefont {Kurz}, \citenamefont {Labiche},
  \citenamefont {{Le Bleis}}, \citenamefont {Lemmon}, \citenamefont {Litvinov},
  \citenamefont {Mahata}, \citenamefont {Maierbeck}, \citenamefont {Movsesyan},
  \citenamefont {Nilsson}, \citenamefont {Nociforo}, \citenamefont {Palit},
  \citenamefont {Paschalis}, \citenamefont {Plag}, \citenamefont {Reifarth},
  \citenamefont {Savran}, \citenamefont {Scheit}, \citenamefont {Simon},
  \citenamefont {S{\"{u}}mmerer}, \citenamefont {Wagner}, \citenamefont
  {Walua}, \citenamefont {Weick},\ and\ \citenamefont {Winkler}}]{Rossi2013}%
  \BibitemOpen
  \bibfield  {author} {\bibinfo {author} {\bibfnamefont {D.~M.}\ \bibnamefont
  {Rossi}}, \bibinfo {author} {\bibfnamefont {P.}~\bibnamefont {Adrich}},
  \bibinfo {author} {\bibfnamefont {F.}~\bibnamefont {Aksouh}}, \bibinfo
  {author} {\bibfnamefont {H.}~\bibnamefont {Alvarez-Pol}}, \bibinfo {author}
  {\bibfnamefont {T.}~\bibnamefont {Aumann}}, \bibinfo {author} {\bibfnamefont
  {J.}~\bibnamefont {Benlliure}}, \bibinfo {author} {\bibfnamefont
  {M.}~\bibnamefont {B{\"{o}}hmer}}, \bibinfo {author} {\bibfnamefont
  {K.}~\bibnamefont {Boretzky}}, \bibinfo {author} {\bibfnamefont
  {E.}~\bibnamefont {Casarejos}}, \bibinfo {author} {\bibfnamefont
  {M.}~\bibnamefont {Chartier}}, \bibinfo {author} {\bibfnamefont
  {A.}~\bibnamefont {Chatillon}}, \bibinfo {author} {\bibfnamefont
  {D.}~\bibnamefont {Cortina-Gil}}, \bibinfo {author} {\bibfnamefont
  {U.}~\bibnamefont {{Datta Pramanik}}}, \bibinfo {author} {\bibfnamefont
  {H.}~\bibnamefont {Emling}}, \bibinfo {author} {\bibfnamefont
  {O.}~\bibnamefont {Ershova}}, \bibinfo {author} {\bibfnamefont
  {B.}~\bibnamefont {Fernandez-Dominguez}}, \bibinfo {author} {\bibfnamefont
  {H.}~\bibnamefont {Geissel}}, \bibinfo {author} {\bibfnamefont
  {M.}~\bibnamefont {Gorska}}, \bibinfo {author} {\bibfnamefont
  {M.}~\bibnamefont {Heil}}, \bibinfo {author} {\bibfnamefont {H.~T.}\
  \bibnamefont {Johansson}}, \bibinfo {author} {\bibfnamefont {A.}~\bibnamefont
  {Junghans}}, \bibinfo {author} {\bibfnamefont {A.}~\bibnamefont
  {Kelic-Heil}}, \bibinfo {author} {\bibfnamefont {O.}~\bibnamefont {Kiselev}},
  \bibinfo {author} {\bibfnamefont {A.}~\bibnamefont {Klimkiewicz}}, \bibinfo
  {author} {\bibfnamefont {J.~V.}\ \bibnamefont {Kratz}}, \bibinfo {author}
  {\bibfnamefont {R.}~\bibnamefont {Kr{\"{u}}cken}}, \bibinfo {author}
  {\bibfnamefont {N.}~\bibnamefont {Kurz}}, \bibinfo {author} {\bibfnamefont
  {M.}~\bibnamefont {Labiche}}, \bibinfo {author} {\bibfnamefont
  {T.}~\bibnamefont {{Le Bleis}}}, \bibinfo {author} {\bibfnamefont
  {R.}~\bibnamefont {Lemmon}}, \bibinfo {author} {\bibfnamefont {Y.~A.}\
  \bibnamefont {Litvinov}}, \bibinfo {author} {\bibfnamefont {K.}~\bibnamefont
  {Mahata}}, \bibinfo {author} {\bibfnamefont {P.}~\bibnamefont {Maierbeck}},
  \bibinfo {author} {\bibfnamefont {A.}~\bibnamefont {Movsesyan}}, \bibinfo
  {author} {\bibfnamefont {T.}~\bibnamefont {Nilsson}}, \bibinfo {author}
  {\bibfnamefont {C.}~\bibnamefont {Nociforo}}, \bibinfo {author}
  {\bibfnamefont {R.}~\bibnamefont {Palit}}, \bibinfo {author} {\bibfnamefont
  {S.}~\bibnamefont {Paschalis}}, \bibinfo {author} {\bibfnamefont
  {R.}~\bibnamefont {Plag}}, \bibinfo {author} {\bibfnamefont {R.}~\bibnamefont
  {Reifarth}}, \bibinfo {author} {\bibfnamefont {D.}~\bibnamefont {Savran}},
  \bibinfo {author} {\bibfnamefont {H.}~\bibnamefont {Scheit}}, \bibinfo
  {author} {\bibfnamefont {H.}~\bibnamefont {Simon}}, \bibinfo {author}
  {\bibfnamefont {K.}~\bibnamefont {S{\"{u}}mmerer}}, \bibinfo {author}
  {\bibfnamefont {A.}~\bibnamefont {Wagner}}, \bibinfo {author} {\bibfnamefont
  {W.}~\bibnamefont {Walua}}, \bibinfo {author} {\bibfnamefont
  {H.}~\bibnamefont {Weick}},\ and\ \bibinfo {author} {\bibfnamefont
  {M.}~\bibnamefont {Winkler}},\ }\href
  {https://doi.org/10.1103/PhysRevLett.111.242503} {\bibfield  {journal}
  {\bibinfo  {journal} {Phys. Rev. Lett.}\ }\textbf {\bibinfo {volume} {111}},\
  \bibinfo {pages} {242503} (\bibinfo {year} {2013})}\BibitemShut {NoStop}%
\bibitem [{\citenamefont {Trzci\ifmmode~\acute{n}\else \'{n}\fi{}ska}\ \emph
  {et~al.}(2001)\citenamefont {Trzci\ifmmode~\acute{n}\else \'{n}\fi{}ska},
  \citenamefont {Jastrz\ifmmode~\mbox{\c{e}}\else \c{e}\fi{}bski},
  \citenamefont {Lubi\ifmmode~\acute{n}\else \'{n}\fi{}ski}, \citenamefont
  {Hartmann}, \citenamefont {Schmidt}, \citenamefont {von Egidy},\ and\
  \citenamefont {K\l{}os}}]{Trzcinska2001a}%
  \BibitemOpen
  \bibfield  {author} {\bibinfo {author} {\bibfnamefont {A.}~\bibnamefont
  {Trzci\ifmmode~\acute{n}\else \'{n}\fi{}ska}}, \bibinfo {author}
  {\bibfnamefont {J.}~\bibnamefont {Jastrz\ifmmode~\mbox{\c{e}}\else
  \c{e}\fi{}bski}}, \bibinfo {author} {\bibfnamefont {P.}~\bibnamefont
  {Lubi\ifmmode~\acute{n}\else \'{n}\fi{}ski}}, \bibinfo {author}
  {\bibfnamefont {F.~J.}\ \bibnamefont {Hartmann}}, \bibinfo {author}
  {\bibfnamefont {R.}~\bibnamefont {Schmidt}}, \bibinfo {author} {\bibfnamefont
  {T.}~\bibnamefont {von Egidy}},\ and\ \bibinfo {author} {\bibfnamefont
  {B.}~\bibnamefont {K\l{}os}},\ }\href
  {https://doi.org/10.1103/PhysRevLett.87.082501} {\bibfield  {journal}
  {\bibinfo  {journal} {Phys. Rev. Lett.}\ }\textbf {\bibinfo {volume} {87}},\
  \bibinfo {pages} {082501} (\bibinfo {year} {2001})}\BibitemShut {NoStop}%
\bibitem [{\citenamefont {Jastrz\c{e}bski}\ \emph {et~al.}(2004)\citenamefont
  {Jastrz\c{e}bski}, \citenamefont {Trzci\'nska}, \citenamefont {Lubi\'nski},
  \citenamefont {K{\l}os}, \citenamefont {Hartmann}, \citenamefont {von
  Egidy},\ and\ \citenamefont {Wycech}}]{JASTRZEBSKI2004}%
  \BibitemOpen
  \bibfield  {author} {\bibinfo {author} {\bibfnamefont {J.}~\bibnamefont
  {Jastrz\c{e}bski}}, \bibinfo {author} {\bibfnamefont {A.}~\bibnamefont
  {Trzci\'nska}}, \bibinfo {author} {\bibfnamefont {P.}~\bibnamefont
  {Lubi\'nski}}, \bibinfo {author} {\bibfnamefont {B.}~\bibnamefont {K{\l}os}},
  \bibinfo {author} {\bibfnamefont {F.~J.}\ \bibnamefont {Hartmann}}, \bibinfo
  {author} {\bibfnamefont {T.}~\bibnamefont {von Egidy}},\ and\ \bibinfo
  {author} {\bibfnamefont {S.}~\bibnamefont {Wycech}},\ }\href
  {https://doi.org/10.1142/S0218301304002168} {\bibfield  {journal} {\bibinfo
  {journal} {Int. J. Mod. Phys. E}\ }\textbf {\bibinfo {volume} {13}},\
  \bibinfo {pages} {343} (\bibinfo {year} {2004})}\BibitemShut {NoStop}%
\bibitem [{\citenamefont {Zhang}\ \emph {et~al.}(2021)\citenamefont {Zhang},
  \citenamefont {Tu}, \citenamefont {Sarriguren}, \citenamefont {Yue},
  \citenamefont {Zeng}, \citenamefont {Sun}, \citenamefont {Wang},
  \citenamefont {Zhang}, \citenamefont {Zhou},\ and\ \citenamefont
  {Litvinov}}]{Zhang2021}%
  \BibitemOpen
  \bibfield  {author} {\bibinfo {author} {\bibfnamefont {J.~T.}\ \bibnamefont
  {Zhang}}, \bibinfo {author} {\bibfnamefont {X.~L.}\ \bibnamefont {Tu}},
  \bibinfo {author} {\bibfnamefont {P.}~\bibnamefont {Sarriguren}}, \bibinfo
  {author} {\bibfnamefont {K.}~\bibnamefont {Yue}}, \bibinfo {author}
  {\bibfnamefont {Q.}~\bibnamefont {Zeng}}, \bibinfo {author} {\bibfnamefont
  {Z.~Y.}\ \bibnamefont {Sun}}, \bibinfo {author} {\bibfnamefont
  {M.}~\bibnamefont {Wang}}, \bibinfo {author} {\bibfnamefont {Y.~H.}\
  \bibnamefont {Zhang}}, \bibinfo {author} {\bibfnamefont {X.~H.}\ \bibnamefont
  {Zhou}},\ and\ \bibinfo {author} {\bibfnamefont {Y.~A.}\ \bibnamefont
  {Litvinov}},\ }\href {https://doi.org/10.1103/PhysRevC.104.034303} {\bibfield
   {journal} {\bibinfo  {journal} {Phys. Rev. C}\ }\textbf {\bibinfo {volume}
  {104}},\ \bibinfo {pages} {034303} (\bibinfo {year} {2021})}\BibitemShut
  {NoStop}%
\bibitem [{\citenamefont {Chabanat}\ \emph {et~al.}(1998)\citenamefont
  {Chabanat}, \citenamefont {Bonche}, \citenamefont {Haensel}, \citenamefont
  {Meyer},\ and\ \citenamefont {Schaeffer}}]{Chabanat1998}%
  \BibitemOpen
  \bibfield  {author} {\bibinfo {author} {\bibfnamefont {E.}~\bibnamefont
  {Chabanat}}, \bibinfo {author} {\bibfnamefont {P.}~\bibnamefont {Bonche}},
  \bibinfo {author} {\bibfnamefont {P.}~\bibnamefont {Haensel}}, \bibinfo
  {author} {\bibfnamefont {J.}~\bibnamefont {Meyer}},\ and\ \bibinfo {author}
  {\bibfnamefont {R.}~\bibnamefont {Schaeffer}},\ }\href
  {https://doi.org/10.1016/S0375-9474(98)00180-8} {\bibfield  {journal}
  {\bibinfo  {journal} {Nucl. Phys. A}\ }\textbf {\bibinfo {volume} {635}},\
  \bibinfo {pages} {231} (\bibinfo {year} {1998})}\BibitemShut {NoStop}%
\bibitem [{\citenamefont {Bespalova}\ and\ \citenamefont
  {Klimochkina}(2017)}]{Bespalova2017}%
  \BibitemOpen
  \bibfield  {author} {\bibinfo {author} {\bibfnamefont {O.~V.}\ \bibnamefont
  {Bespalova}}\ and\ \bibinfo {author} {\bibfnamefont {A.~A.}\ \bibnamefont
  {Klimochkina}},\ }\href {https://doi.org/10.1134/S1063778817050039}
  {\bibfield  {journal} {\bibinfo  {journal} {Phys. At. Nucl.}\ }\textbf
  {\bibinfo {volume} {80}},\ \bibinfo {pages} {919} (\bibinfo {year}
  {2017})}\BibitemShut {NoStop}%
\bibitem [{\citenamefont {Thiel}\ \emph {et~al.}(2019)\citenamefont {Thiel},
  \citenamefont {Sfienti}, \citenamefont {Piekarewicz}, \citenamefont
  {Horowitz},\ and\ \citenamefont {Vanderhaeghen}}]{Thiel2019}%
  \BibitemOpen
  \bibfield  {author} {\bibinfo {author} {\bibfnamefont {M.}~\bibnamefont
  {Thiel}}, \bibinfo {author} {\bibfnamefont {C.}~\bibnamefont {Sfienti}},
  \bibinfo {author} {\bibfnamefont {J.}~\bibnamefont {Piekarewicz}}, \bibinfo
  {author} {\bibfnamefont {C.~J.}\ \bibnamefont {Horowitz}},\ and\ \bibinfo
  {author} {\bibfnamefont {M.}~\bibnamefont {Vanderhaeghen}},\ }\href
  {https://doi.org/10.1088/1361-6471/ab2c6d} {\bibfield  {journal} {\bibinfo
  {journal} {J. Phys. G: Nucl. Part. Phys}\ }\textbf {\bibinfo {volume} {46}},\
  \bibinfo {pages} {093003} (\bibinfo {year} {2019})}\BibitemShut {NoStop}%
\bibitem [{\citenamefont {Bellicard}\ \emph {et~al.}(1967)\citenamefont
  {Bellicard}, \citenamefont {Bounin}, \citenamefont {Frosch}, \citenamefont
  {Hofstadter}, \citenamefont {McCarthy}, \citenamefont {Uhrhane},
  \citenamefont {Yearian}, \citenamefont {Clark}, \citenamefont {Herman},\ and\
  \citenamefont {Ravenhall}}]{Bellicard67}%
  \BibitemOpen
  \bibfield  {author} {\bibinfo {author} {\bibfnamefont {J.~B.}\ \bibnamefont
  {Bellicard}}, \bibinfo {author} {\bibfnamefont {P.}~\bibnamefont {Bounin}},
  \bibinfo {author} {\bibfnamefont {R.~F.}\ \bibnamefont {Frosch}}, \bibinfo
  {author} {\bibfnamefont {R.}~\bibnamefont {Hofstadter}}, \bibinfo {author}
  {\bibfnamefont {J.~S.}\ \bibnamefont {McCarthy}}, \bibinfo {author}
  {\bibfnamefont {F.~J.}\ \bibnamefont {Uhrhane}}, \bibinfo {author}
  {\bibfnamefont {M.~R.}\ \bibnamefont {Yearian}}, \bibinfo {author}
  {\bibfnamefont {B.~C.}\ \bibnamefont {Clark}}, \bibinfo {author}
  {\bibfnamefont {R.}~\bibnamefont {Herman}},\ and\ \bibinfo {author}
  {\bibfnamefont {D.~G.}\ \bibnamefont {Ravenhall}},\ }\href
  {https://doi.org/10.1103/PhysRevLett.19.527} {\bibfield  {journal} {\bibinfo
  {journal} {Phys. Rev. Lett.}\ }\textbf {\bibinfo {volume} {19}},\ \bibinfo
  {pages} {527} (\bibinfo {year} {1967})}\BibitemShut {NoStop}%
\bibitem [{\citenamefont {Migdal}(1983)}]{Migdal1983}%
  \BibitemOpen
  \bibfield  {author} {\bibinfo {author} {\bibfnamefont {A.}~\bibnamefont
  {Migdal}},\ }\href@noop {} {\emph {\bibinfo {title} {{The Theory of Finite
  Fermi-Systems and Properties of Atomic Nuclei}}}},\ \bibinfo {edition} {2nd}\
  ed.\ (\bibinfo  {publisher} {Nauka},\ \bibinfo {address} {Moscow},\ \bibinfo
  {year} {1983})\BibitemShut {NoStop}%
\bibitem [{\citenamefont {Speth}\ \emph {et~al.}(2014)\citenamefont {Speth},
  \citenamefont {Krewald}, \citenamefont {Gru¨mmer}, \citenamefont {Reinhard},
  \citenamefont {Lyutorovich},\ and\ \citenamefont {Tselyaev}}]{Speth2014}%
  \BibitemOpen
  \bibfield  {author} {\bibinfo {author} {\bibfnamefont {J.}~\bibnamefont
  {Speth}}, \bibinfo {author} {\bibfnamefont {S.}~\bibnamefont {Krewald}},
  \bibinfo {author} {\bibfnamefont {F.}~\bibnamefont {Gru¨mmer}}, \bibinfo
  {author} {\bibfnamefont {P.~G.}\ \bibnamefont {Reinhard}}, \bibinfo {author}
  {\bibfnamefont {N.}~\bibnamefont {Lyutorovich}},\ and\ \bibinfo {author}
  {\bibfnamefont {V.}~\bibnamefont {Tselyaev}},\ }\href
  {https://doi.org/10.1016/j.nuclphysa.2014.03.023} {\bibfield  {journal}
  {\bibinfo  {journal} {Nucl. Phys. A}\ }\textbf {\bibinfo {volume} {928}},\
  \bibinfo {pages} {17} (\bibinfo {year} {2014})}\BibitemShut {NoStop}%
\bibitem [{\citenamefont {Simonov}\ \emph {et~al.}(2022)\citenamefont
  {Simonov}, \citenamefont {Karpov},\ and\ \citenamefont
  {Tretyakova}}]{Simonov2022}%
  \BibitemOpen
  \bibfield  {author} {\bibinfo {author} {\bibfnamefont {M.~V.}\ \bibnamefont
  {Simonov}}, \bibinfo {author} {\bibfnamefont {A.~V.}\ \bibnamefont
  {Karpov}},\ and\ \bibinfo {author} {\bibfnamefont {T.~Y.}\ \bibnamefont
  {Tretyakova}},\ }\href
  {https://link.springer.com/article/10.3103/S1062873822080202} {\bibfield
  {journal} {\bibinfo  {journal} {Bull. Russ. Acad. Sci. Phys.}\ }\textbf
  {\bibinfo {volume} {86}},\ \bibinfo {pages} {931} (\bibinfo {year}
  {2022})}\BibitemShut {NoStop}%
\bibitem [{\citenamefont {B{\l}ocki}\ \emph {et~al.}(1977)\citenamefont
  {B{\l}ocki}, \citenamefont {Randrup}, \citenamefont {{\'{S}}wi\c{a}tecki},\
  and\ \citenamefont {Tsang}}]{Blocki1977}%
  \BibitemOpen
  \bibfield  {author} {\bibinfo {author} {\bibfnamefont {J.}~\bibnamefont
  {B{\l}ocki}}, \bibinfo {author} {\bibfnamefont {J.}~\bibnamefont {Randrup}},
  \bibinfo {author} {\bibfnamefont {W.~J.}\ \bibnamefont
  {{\'{S}}wi\c{a}tecki}},\ and\ \bibinfo {author} {\bibfnamefont {C.~F.}\
  \bibnamefont {Tsang}},\ }\href {https://doi.org/10.1016/0003-4916(77)90249-4}
  {\bibfield  {journal} {\bibinfo  {journal} {Ann. Phys. (N. Y).}\ }\textbf
  {\bibinfo {volume} {105}},\ \bibinfo {pages} {427} (\bibinfo {year}
  {1977})}\BibitemShut {NoStop}%
\end{thebibliography}
\end{document}